\documentclass[11pt,a4paper]{article}
\usepackage{jheppub}
\usepackage{latexsym}

\newcommand{\be}{\begin{equation}}
\newcommand{\ee}{\end{equation}}

\def\cN{{\cal N}}
\def\dfrac{\displaystyle \frac }

\newcommand{\bea}{\begin{eqnarray}}
\newcommand{\eea}{\end{eqnarray}}
\newcommand{\ba}{\begin{array}}
\newcommand{\ea}{\end{array}}
\newcommand{\nn}{\nonumber}

\title{Superconformal $\cN=3$ SYM Low-Energy \\ Effective Action}

\author[a]{I.L. Buchbinder,}
\author[b]{E.A. Ivanov,}
\author[c,1]{I.B. Samsonov%
\note{On leave from Tomsk Polytechnic University, 634050 Tomsk,
Russia.}}
\author[b]{and B.M. Zupnik}

\affiliation[a]{Departamento de Fisica, UFJF, Juiz de Fora, MG,
Brazil
 and \\
 Department of Theoretical Physics, Tomsk State Pedagogical
 University, Tomsk 634061, Russia \footnote{Permanent address.}}

\affiliation[b]{Bogolubov Laboratory of Theoretical Physics, Joint
Institute for Nuclear Research, Dubna, 141980 Moscow Region,
Russia}

\affiliation[c]{INFN, Sezione di Padova, via F. Marzolo 8, 35131
Padova, Italy}

\emailAdd{joseph@tspu.edu.ru}
\emailAdd{eivanov@theor.jinr.ru}
\emailAdd{samsonov@mph.phtd.tpu.ru}
\emailAdd{zupnik@theor.jinr.ru}

\abstract{We construct a manifestly $\cN=3$ supersymmetric low-energy
effective action of $\cN=3$ super Yang-Mills theory. The effective
action is written in the $\cN=3$ harmonic superspace and respects
the full $\cN=3$ superconformal symmetry. On mass shell this action is
responsible for the four-derivative terms in the $\cN=4$ SYM
effective action, such as $F^4/X^4$ and its supersymmetric
completions, while off shell it involves also higher-derivative
terms. For constant Maxwell and scalar fields its bosonic part  coincides, up to the $F^6/X^8$ order,
with the bosonic part of the D3 brane action in the $AdS_5\times S^5$ background. We
also argue that in the sector of scalar fields it involves the correctly normalized Wess-Zumino term with
the implicit SU(3) symmetry.}


\arxivnumber{1111.4145}

\begin{document}
\makeatletter
\gdef\@fpheader{}
\makeatother
\maketitle
\section{Introduction}

It is well known that the superfield formulations of supersymmetric
field theories, with the maximal number of the underlying
supersymmetries being manifest and off-shell, are extremely useful
for studying quantum aspects of these theories. In many cases, such
formulations not only drastically reduce  the amount of perturbative
calculations, but also allow one to make certain conjectures about a
possible structure of the final results prior to any calculation.
The $\cN=1$ superspace \cite{BK-book} is natural for $\cN=1$, $d=4$
supersymmetric models, while the adequate superfield approach to
$\cN=2$, $d=4$ theories is offered by $\cN=2$ harmonic superspace
\cite{hss-book}.
As for the renowned $\cN=4$ SYM theory, no appropriate formulation of it  in
terms of unconstrained off-shell $\cN=4$ superfields is known to date.

On shell, the $\cN=4$ SYM theory is equivalent to the $\cN=3$ SYM theory (see \cite{hss-book} and refs.\ therein).
The latter  possesses an unconstrained superfield formulation in $\cN=3$ harmonic superspace \cite{GIKOS1,GIKOS2}, such
that three out of four supersymmetries  of the original theory are manifest and off-shell within this framework.
This approach proved to be very fruitful for establishing quantum finiteness of $\cN=3$ SYM theory \cite{DM}, as well as
for constructing the $\cN=3$ supersymmetric Born-Infeld theory
\cite{IZ}.\footnote{A possible scale-invariant generalization of the $\cN=3$ Born-Infeld theory was discussed in \cite{BISZ}.}
The basic goal of the present paper is to provide an evidence that the $\cN=3$ harmonic superspace approach is also
useful for studying the low-energy effective actions in the $\cN=3$ and $\cN=4$ SYM theories.

It is known that the leading terms in the $\cN=4$ SYM effective action in the $\cN=2$ superfield formulation (which manifests only
two out of four supersymmetries) are described by the non-holomorphic
potential \cite{DS,S,deWit:1996,non-hol1,non-hol2,non-hol3,non-hol4,non-hol5,review}.
The hypermultiplet completion of the non-holomorphic potential,
such that it ensures the on-shell $\cN=4$ supersymmetry, was found in
\cite{BuIv} (and further elaborated on in refs.
\cite{BIP,BBP,BuPl}). The full action contains a scale-invariant
and SU(4) symmetric $F^4/X^4$ term, as well as some other terms
related to this leading one by $\cN=4$ supersymmetry. Here
$F_{mn}$ is the Maxwell field strength and $X^2$ is the square of
SU(4) invariant norm of scalar fields.

In the present paper we develop the $\cN=3$ harmonic superspace description of the leading terms in the
$\cN=4$ SYM effective action to the order $F^4/X^4$. We seek this action as an integral over the analytic
subspace of the $\cN=3$ harmonic superspace, with the Lagrangian density  being a local functional of the
analytic superfield strengths without derivatives on them. We show
that the requirements of the scale and $\gamma_5$ invariance uniquely fix
the form of this functional. We check that the action constructed respects the full SU(2,2$|$3) superconformal symmetry
and, in components, yields the $F^4/X^4$ term, where $X^2=\varphi^i\bar\varphi_i$ is the bilinear SU(3)
(and in fact SU(4)) invariant of the involved scalar fields.

We stress that the obtained action is essentially defined on the
Coulomb branch of the theory, when the scalar fields acquire
non-vanishing vevs, $c^i=\langle\varphi^i \rangle\ne0$. These
constants $c^i$ explicitly appear in the effective action, so that the
effective Lagrangian is singular at $c^i=0$. However, we show that
the action is in fact independent of any particular choice of $c^i$,
$c^i\ne0$. This is entirely analogous to what happens in the  $\cN=2$
harmonic superspace formulation of the $\cN=2$ improved tensor
multiplet model given in \cite{GIO}. It was emphasized there that the
presence of such constants in the action has a topological origin.
In accord with this interpretation, the low-energy $\cN=4$
SYM effective action contains a topological term given by the
Wess-Zumino action for the scalar fields \cite{TZ,Intriligator}.
Therefore the presence of such constants in the effective
Lagrangian is not surprising.

One of the advantages of the $\cN=3$ superspace formulation is the
possibility to go off shell due to the existence of unconstrained
gauge prepotentials. Varying with respect to these prepotentials, we obtain the
effective equations of motion corresponding to the effective action.
Like in the non-scale-invariant $\cN=3$ Born-Infeld
theory \cite{IZ}, elimination of some of the auxiliary fields from the
effective equations of motion allows one to reproduce not only $F^4/X^4$ term,
but also the  $F^6/X^8$ term in the effective action, which precisely
matches with that appearing in the component expansion of the conformally-invariant Born-Infeld
action. Therefore, the  effective action obtained reproduces the
worldvolume action of D3 brane on the $AdS_5\times S^5$ background up to the order
$F^6/X^8$.

The paper is organized as follows. Section 2 contains a brief
summary of the basic ingredients of the $\cN=3$ harmonic
superspace formalism. In particular, we give the representation of
the $\cN=3$ superconformal group  SU(2,2$|$3) on the $\cN=3$ SYM
superfield strengths. In Section 3, employing the scale and
$\gamma_5$ invariance, we derive the $\cN=3$ SYM low-energy
effective action and then show its invariance under the full
$\cN=3$ superconformal group. In Section 4 we derive the $F^4/X^4$
and $F^6/X^8$ component terms from the superfield action. The last
Section is devoted to discussing some open problems deserving
further study. In Appendices A and B we collect some technical
details concerning the derivation of the superfield action and
calculation of SU(3) harmonic integrals. A possible
four-dimensional representation of the Wess-Zumino term with
manifest SU(3) symmetry is discussed in the Appendix C. In
Appendix D we demonstrate that the $\cN=3$ superfield effective
action proposed can be used for studying the effective superfield
equations of motion.

Throughout the paper we follow the $\cN=3$ superspace
conventions employed in \cite{BISZ} and \cite{IZ}.

\section{$\cN=3$ SYM setup}
\subsection{Superfield strengths in $\cN=3$ harmonic superspace}

The standard $\cN=3$ superspace is parametrized by the coordinates
$z^M=(x^m,\theta_i^\alpha,\bar\theta^{i\dot\alpha})\,$, where
$i=1,2,3$ is the SU(3) triplet index. Following \cite{GIKOS1,GIKOS2},
we introduce the SU(3) harmonic variables
$u^I_i=(u^1_i, u^2_i, u^3_i)$ and their conjugates,
$\bar u_I^i=(\bar u_1^i, \bar u_2^i,\bar u_3^i)$, with the
properties
\be
u^I_i \bar u^i_J=\delta^I_J\,,\quad
u^I_i \bar u^j_I=\delta_i^j\,,\quad
\varepsilon^{ijk}u^1_i u^2_j u^3_k=1\,.
\ee
These defining relations are the orthogonality and
completeness conditions. The harmonic variables  allow one to convert the small indices $i,j,\ldots$ on
which the R-symmetry SU(3) group is linearly realized, into the capital indices, $I,J,\ldots$, which are inert under SU(3).
For instance, we will make use of the projected Grassmann variables,
$\theta_I^\alpha=\theta_i^\alpha \bar u^i_I$,
$\bar\theta^{I\dot\alpha}=\bar\theta^{i\dot\alpha} u^I_i$. Some of
these projected Grassmann variables parametrize the analytic
subspace,
\be
\{\zeta_A, u \}=\{x_A^{m},\theta_2^\alpha,\theta_3^\alpha,
\bar\theta^{1\dot\alpha},\bar\theta^{2\dot\alpha},u \}\,,\qquad
x_A^{m}=x^m-i\theta_1\sigma^m\bar\theta^1+i\theta_3\sigma^m\bar\theta^3\,.
\label{anal-coord}
\ee
The analytic superspace (\ref{anal-coord}) is closed
under the $\cN=3$ supersymmetry \cite{GIKOS1,GIKOS2}, and, hence, plays a role similar
to that of usual chiral subspace in the $\cN=1$ superspace \cite{BK-book} and of the $\cN=2$ harmonic
analytic superspace \cite{hss-book}.

The harmonic projections of the covariant spinor derivatives\footnote{
We use the following rules of converting the vector and bi-spinorial indices into
each other,
$x_{\alpha\dot\alpha}=(\sigma^m)_{\alpha\dot\alpha}x_m$,
$x_m=\frac12(\tilde\sigma_m)^{\dot\alpha\alpha}x_{\alpha\dot\alpha}$,
$\partial_{\alpha\dot\alpha}=\frac12(\sigma^m)_{\alpha\dot\alpha}\partial_m$,
$\partial_m=(\tilde\sigma_m)^{\dot\alpha\alpha}\partial_{\alpha\dot\alpha}$.},
\be
D^i_\alpha=\frac{\partial}{\partial\theta_i^\alpha}+2i\bar\theta^{i\dot\alpha}
\frac{\partial}{\partial x^{\alpha\dot\alpha}}\,,\qquad
\bar D_{i\dot\alpha}=-\frac{\partial}{\partial\bar\theta^{i\dot\alpha}}-
2i\theta_i^\alpha\frac{\partial}{\partial x^{\alpha\dot\alpha}}\,,
\ee
are given by $D^I_\alpha=D^i_\alpha  u^I_i$ and
$\bar D_{I\dot\alpha}=\bar D_{i\dot\alpha}\bar u_I^i$. It is
important that in the analytic coordinates
(\ref{anal-coord}) two of these six derivatives become short,
\be
D^1_\alpha=\frac\partial{\partial\theta_1^\alpha}\,,\qquad
\bar
D_{3\dot\alpha}=-\frac\partial{\partial\bar\theta^{3\dot\alpha}}\,,
\label{anal-deriv}
\ee
thus demonstrating that the $\cN=3$ analytic superfields (i.e.\ those living on the analytic superspace (\ref{anal-coord}))
can be covariantly defined by the Grassmann Cauchy-Riemann
conditions,
\be
D^1_\alpha \Phi(z, u) =\bar  D_{3\dot\alpha} \Phi(z, u) = 0\quad
 \Rightarrow \quad\Phi(z, u) = \hat\Phi(\zeta_A, u)\,.
\ee

The explicit expressions for the other four derivatives in the analytic basis can be
found in the appendix of our previous paper \cite{BISZ}, where the
harmonic derivatives $D^I_J$ are also written down.
Among these harmonic derivatives, $D^1_2$, $D^2_3$ and $D^1_3$
commute with (\ref{anal-deriv}) and so preserve the analyticity, while the remaining three $D^2_1$, $D^3_2$, $D^3_1$
do not. These six harmonic derivatives, together with the U(1)
charges $S_1$ and $S_2$, form an su(3) algebra \cite{AFSZ}.

The conventional $\cN=3$ SYM superfield strengths in the standard $\cN=3$ superspace
are described by the antisymmetric SU(3) tensor superfields
$W^{ij}=-W^{ji}$. In the linearized approximation, these superfields obey the constraints \cite{HST},
\be
D^i_\alpha W_{jl}=\frac12(\delta^i_j D^k_\alpha W_{kl}-\delta^i_l D^k_\alpha W_{kj})
\,,\qquad
\bar D_{i\dot\alpha}W_{jk}+\bar D_{j\dot\alpha}W_{ik}=0\,,
\label{constr}
\ee
which eliminate all non-physical components in these superfields and
put the physical ones on shell. Projecting these superfield
strengths on the harmonic variables, we obtain the following six
superfields,
\bea
&&
\bar W^{12}=u^1_i u^2_j\bar  W^{ij}\,,\quad
\bar W^{23}=u^2_i u^3_j\bar  W^{ij}\,,\quad
\bar W^{13}=u^1_i u^3_j\bar  W^{ij}\,,\quad
\nn\\&&
W_{12}=\bar u_1^i \bar u_2^j W_{ij}\,,\quad
W_{23}=\bar u_2^i \bar u_3^j W_{ij}\,,\quad
W_{13}=\bar u_1^i \bar u_3^j W_{ij}\,.
\label{W_}
\eea

It is straightforward to find the harmonic projections of the
constraints (\ref{constr}), which gives rise to a number of differential
relations among the superfields (\ref{W_}).
Consider, for instance, $\bar W^{12}$ and $W_{23}$. They
obey the following (on-shell) constraints \cite{AFSZ}\footnote{The constraints (\ref{anal})--(\ref{linearity}) can
also be derived by quantizing a massless superparticle moving in the $\cN=3$
harmonic superspace \cite{BS08}.}:
\begin{itemize}
\item[(i)] First-order analyticity constraints,
\bea
&&
D^1_\alpha \bar W^{12}=D^2_\alpha \bar W^{12}=\bar D_{3\dot\alpha}
\bar W^{12}=0\,,\nn\\&&
D^1_\alpha W_{23}=\bar D_{2\dot\alpha}W_{23}=\bar
D_{3\dot\alpha}W_{23}=0\,;
\label{anal}
\eea
\item[(ii)] First-order harmonic shortness constraints,
\bea
&&
D^2_1\bar W^{12} =
D^1_2 \bar W^{12}=D^2_3 \bar W^{12}=D^1_3 \bar W^{12}=0\,,\nn\\&&
D^1_2 W_{23}=D^2_3 W_{23}=D^1_3 W_{23}=D^3_2 W_{23}=0\,;
\label{harm-short}
\eea
\item[(iii)] Second-order Grassmann linearity constraints,
\bea
&&(D^3)^2 \bar W^{12}=(\bar D_1)^2 \bar W^{12}=(\bar D_2)^2
\bar W^{12}=(\bar D_1 \bar D_2)\bar W^{12}=0\,,\nn\\&&
(D^2)^2W_{23}=(D^3)^2 W_{23}=(D^2 D^3)W_{23}=(\bar D_1)^2
W_{23}=0\,.
\label{linearity}
\eea
\end{itemize}
Altogether, the constraints (\ref{anal}), (\ref{harm-short}) and
(\ref{linearity}) kill all non-physical (auxiliary) field
components in $\bar W^{12}$ and $W_{23}$ and put the physical ones on shell\footnote{Besides eqs. (\ref{harm-short}),
the original constraints (\ref{constr}) imply some other relations of the first order in spinor derivatives, connecting $\bar W^{12}$ and $W_{23}$
with the remaining harmonic projections of $W_{kl}$ and $\bar W^{kl}$. These extra constraints can be used to deduce
the second-order constraints (\ref{linearity}) which, together with (\ref{anal}) and (\ref{harm-short}), form a closed set
of the harmonic superspace constraints on $\bar W^{12}$ and $W_{23}$ \cite{BS08}.}.
As a result, the superfield strengths
$\bar W^{12}$ and $W_{23}$ have the following component structure in the analytic
coordinates (\ref{anal-coord}),
\bea
\label{W-comp}
W_{23}&=&\varphi^1
+2i\theta_2^\alpha\bar\theta^{2\dot\alpha}\partial_{\alpha\dot\alpha}\varphi^1
-4i\theta_2^\alpha\bar\theta^{1\dot\alpha}\partial_{\alpha\dot\alpha}\varphi^2
-4i\theta_3^\alpha\bar\theta^{1\dot\alpha}\partial_{\alpha\dot\alpha}\varphi^3
\nn\\&&
+\,4i\theta_2^\alpha\theta_3^\beta F_{\alpha\beta}
+\bar\theta^{1\dot\alpha}\bar\lambda_{\dot\alpha}
+\theta_2^\alpha\lambda_{3\alpha}-\theta_3^\alpha\lambda_{2\alpha}
\nn\\&&
+\,2i\theta_2^\alpha\bar\theta^{2\dot\alpha}\bar\theta^{1\dot\beta}
 \partial_{\alpha\dot\alpha}\bar\lambda_{\dot\beta}
+2i\theta_2^\beta\theta_3^\alpha \bar\theta^{2\dot\alpha}
 \partial_{\alpha\dot\alpha}\lambda_{2\beta}
+4i\theta_2^\beta\theta_3^\alpha\bar\theta^{1\dot\alpha}
 \partial_{\alpha\dot\alpha}\lambda_{1\beta}
 \nn\\&&
+\,8\theta_2^\alpha\theta_3^\beta\bar\theta^{1\dot\alpha}\bar\theta^{2\dot\beta}
\partial_{\alpha\dot\alpha}\partial_{\beta\dot\beta}\varphi^3\,,
\nn\\
\bar W^{12}&=&\bar\varphi_3
-2i\theta_2^\alpha\bar\theta^{2\dot\alpha}\partial_{\alpha\dot\alpha}\bar\varphi_3
+4i\theta_3^\alpha\bar\theta^{1\dot\alpha}\partial_{\alpha\dot\alpha}\bar\varphi_1
+4i\theta_3^\alpha\bar\theta^{2\dot\alpha}\partial_{\alpha\dot\alpha}\bar\varphi_2
\nn\\&&
+\,4i\bar\theta^{1\dot\alpha}\bar\theta^{2\dot\beta}\bar F_{\dot\alpha\dot\beta}
+\theta_3^\alpha\lambda_\alpha
-\bar\theta^{2\dot\alpha}\bar\lambda^1_{\dot\alpha}
+\bar\theta^{1\dot\alpha}\bar\lambda^2_{\dot\alpha}
\nn\\&&
+\,2i\theta_2^\alpha\theta_3^\beta\bar\theta^{2\dot\alpha}
 \partial_{\alpha\dot\alpha}\lambda_\beta
+2i\bar\theta^{1\dot\alpha}\bar\theta^{2\dot\beta}\theta_2^\alpha
 \partial_{\alpha\dot\alpha}\bar\lambda^2_{\dot\beta}
+4i\bar\theta^{1\dot\alpha}\bar\theta^{2\dot\beta}\theta_3^\alpha
 \partial_{\alpha\dot\alpha}\bar\lambda^3_{\dot\beta}
\nn\\&&
+\,8\bar\theta^{1\dot\alpha}\bar\theta^{2\dot\beta}
  \theta_2^\alpha\theta_3^\beta
  \partial_{\alpha\dot\alpha}\partial_{\beta\dot\beta}\bar\varphi_1
  \,.
  \label{W}
\eea
Here
\be
\varphi^I=u^I_i \varphi^i\,,\qquad
\bar\varphi_I=\bar u_I^i \bar\varphi_i\,,
\ee
and $\varphi^i$ is a triplet of physical scalars, $\square\varphi^i=0$.
The four spinor fields are comprised by the SU(3) singlet
$\lambda_\alpha$ and the triplet $\lambda_{I\alpha}=\bar
u_I^i\lambda_{i\alpha}$ which obey free equations of motion,
$\partial^{\alpha\dot\alpha}\lambda_\alpha=\partial^{\alpha\dot\alpha}\lambda_{i\alpha}=0$.
The fields $F_{\alpha\beta}=F_{(\alpha\beta)}$ and
$\bar F_{\dot\alpha\dot\beta}=\bar F_{(\dot\alpha\dot\beta)}$ are
spinorial components of the Maxwell field strength $F_{mn}=\partial_m A_n-\partial_n
A_m$, $\partial^m F_{mn}=0$.

The crucial feature of the $\cN=3$  harmonic superspace approach
is that one can relax some of the constraints (\ref{anal}),
(\ref{harm-short}), (\ref{linearity}) and express the superfield
strengths in terms of unconstrained off-shell gauge superfield
potentials \cite{IZ}. Consider the analytic superfields $V^1_2$
and $V^2_3$, $D^1_\alpha(V^1_2,\;V^2_3)=\bar
D_{3\dot\alpha}(V^1_2,\;V^2_3)=0\,$. They possess the following gauge
transformations
\be
\delta V^1_2=iD^1_2 \lambda\,,\qquad
\delta V^2_3=iD^2_3 \lambda\,,
\ee
with $\lambda$ being an analytic gauge superfield parameter.
Using these superfields, one constructs the non-analytic gauge
potentials $V^2_1$ and $V^3_2$ as solutions of the
zero-curvature equations \cite{IZ},
\be
D^1_2 V^2_1=D^2_1 V^1_2\,,\qquad
D^2_3 V^3_2=D^3_2 V^2_3\,.
\label{zero}
\ee
Finally, the gauge-invariant superfield strengths $\bar W^{12}$ and $W_{23}$ can be
expressed in terms of $V^2_1$ and $V^3_2$ as
\be
\bar W^{12}=-\frac14 D^{1\alpha}D^1_\alpha V^2_1\,,\qquad
W_{23}=\frac14 \bar D_{3\dot\alpha}\bar D_3^{\dot\alpha}V^3_2\,.
\label{W-anal}
\ee

It should be pointed out that the analyticity constraints
(\ref{anal}) are valid off shell while the other constraints
(\ref{harm-short}) and (\ref{linearity}) put the superfield
strengths on shell, except for the equations
$D^1_2 \bar W^{12}=0$ and $D^2_3 W_{23}=0$ which are also satisfied
off shell.

\subsection{Superconformal transformations in $\cN=3$ HSS}

The $\cN=3$ superconformal group SU(2,2$|$3), besides the $\cN=3$ super Poincar\'e transformations, contains
dilatation (with the parameter $a$), $\gamma_5$-transformation (with the parameter $b$),
conformal boosts (with the parameters $k_{\alpha\dot\alpha}$), conformal
supersymmetry (with the parameters $\eta^i_\alpha$, $\bar\eta_{i\dot\beta}$) and SU(3)
R-symmetry transformations (with the parameters $\lambda_i^j$, $\overline{\lambda_i^j}=-\lambda_j^i$,
$\lambda_i^i=0$). The realization of this supergroup on the analytic coordinates
(\ref{anal}) was found in \cite{GIO-N3},
\bea
\delta_{\rm sc} x_A^{\alpha\dot\alpha}&=&ax_A^{\alpha\dot\alpha}+
 k_{\beta\dot\beta}x_A^{\alpha\dot\beta}x_A^{\beta\dot\alpha}-
 4k_{\beta\dot\beta}\theta_2^\beta\bar\theta^{2\dot\alpha}
  \theta_2^\alpha\bar\theta^{2\dot\beta}+
  4ix_A^{\alpha\dot\beta}\bar\theta^{1\dot\alpha}\bar
  u_1^i\bar\eta_{i\dot\beta}\nonumber\\
&& +\,2ix_{A-}^{\alpha\dot\beta}\bar\theta^{2\dot\alpha}
  \bar u_2^i \bar\eta_{i\dot\beta}+
  4ix_A^{\beta\dot\alpha}\theta_3^\alpha u^3_i\eta^i_\beta+
  2ix_{A+}^{\beta\dot\alpha}\theta_2^\alpha u^2_i\eta^i_\beta\nonumber\\
&& -\,4i\lambda_i^j\theta_3^\alpha\bar\theta^{1\dot\alpha}u^3_j\bar u_1^i-
  2i\lambda_i^j\theta_2^\alpha\bar\theta^{1\dot\alpha}u^2_j\bar u_1^i-
  2i\lambda_i^j\theta_3^\alpha\bar\theta^{2\dot\alpha}u^3_j\bar
  u_2^i\,,
\nn\\
\delta_{\rm sc}\theta_2^\alpha&=&(a/2+ib)\theta_2^\alpha+
  k_{\beta\dot\beta}x_{A+}^{\alpha\dot\beta}\theta_2^\beta-
  4i(\theta_2^\alpha u^2_i+\theta_3^\alpha u^3_i)\theta_2^\beta\eta^i_\beta
  \nonumber\\
&& +\,x_{A+}^{\alpha\dot\beta}\bar u_2^i\bar\eta_{\dot\beta i}+
   \lambda_i^j(\theta_2^\alpha u^2_j+\theta_3^\alpha u^3_j)\bar
   u_2^i\,,
   \nn\\
\delta_{\rm sc}\theta_3^\alpha&=&(a/2+ib)\theta_3^\alpha+
  k_{\beta\dot\beta}x_{A-}^{\alpha\dot\beta}\theta_3^\beta-
  4i\theta_3^\alpha\theta_3^\beta u^3_i\eta^i_\beta
 +x_{A-}^{\alpha\dot\beta}\bar u_3^i\bar\eta_{\dot\beta i}+
   \lambda_i^j\theta_3^\alpha u^3_j\bar u_3^i\,,
\nn\\
\delta_{\rm sc}\bar\theta^{1\dot\alpha}&=&(a/2-ib)\bar\theta^{1\dot\alpha}+
 k_{\beta\dot\beta}x_{A+}^{\beta\dot\alpha}\bar\theta^{1\dot\beta}+
 4i\bar\theta^{1\dot\beta}\bar\theta^{1\dot\alpha}\bar
  u_1^i\bar\eta_{\dot\beta i}+x_{A+}^{\beta\dot\alpha}u^1_i\eta^i_\beta-
  \lambda_i^j\bar\theta^{1\dot\alpha}\bar u_1^i u^1_j\,,
\nn  \\
\delta_{\rm sc}\bar\theta^{2\dot\alpha}&=&(a/2-ib)\bar\theta^{2\dot\alpha}+
 k_{\beta\dot\beta}x_{A-}^{\beta\dot\alpha}\bar\theta^{2\dot\beta}+
 4i\bar\theta^{2\dot\beta}(\bar\theta^{1\dot\alpha}\bar u_1^i+
 \bar\theta^{2\dot\alpha}\bar u_2^i)\bar\eta_{\dot\beta i}
 \nonumber\\
&&  +\,x_{A-}^{\beta\dot\alpha}u^2_i\eta^i_\beta-
  \lambda_i^j(\bar\theta^{1\dot\alpha}\bar u_1^i +
  \bar\theta^{2\dot\alpha}\bar u_2^i)u^2_j\,,
  \label{dxA}
\end{eqnarray}
where $x_{A\pm}^{\alpha\dot\alpha}=x_{A}^{\alpha\dot\alpha}\pm
2i\theta_2^\alpha\bar\theta^{2\dot\alpha}$.
For preserving the $\cN=3$ harmonic analyticity, the harmonic variables should transform
according to the rules,
\be
\begin{array}{ll}
\delta_{\rm sc} u^1_i=u^2_i\lambda^1_2+u^3_i\lambda^1_3\,,\quad&
 \delta_{\rm sc}\bar u_1^i=0\,,\\
\delta_{\rm sc} u_i^2=u_i^3\lambda^2_3\,,&
 \delta_{\rm sc}\bar u_2^i=-\bar u^i_1\lambda^1_2\,,\\
\delta_{\rm sc} u^3_i=0\,,&
 \delta_{\rm sc}\bar u_3^i=-\bar u_2^i\lambda^2_3-\bar u_1^i\lambda^1_3\,,
\end{array}
\label{du}
\ee
where
\be
\lambda^I_J=-4ik_{\beta\dot\beta}\theta_J^\beta\bar\theta^{I\dot\beta}-
 4i(\bar\eta_{\dot\beta i}\bar\theta^{I\dot\beta}\bar u_J^i+
  \theta_J^\beta\eta^i_\beta u^I_i)+u^I_i\bar u_J^j\lambda^i_j\,.
\label{lambda}
\ee

In this paper we will use the so-called passive form of superconformal
transformations of superfields, when the variation is taken at
different points, e.g., $\delta_{\rm sc} W \simeq W'(x')-W(x)$. In such an approach,
not only the superfields but also their derivatives, as well as the superspace measures, should
be varied while computing the superconformal transformations of the
superfield actions.

It is known \cite{hss-book} that
the analytic measure $d\zeta(^{33}_{11})du$ is invariant under
(\ref{dxA}) and (\ref{du}),
\be
{\rm Ber}\,\frac{\partial(x_A',\theta',u')}{\partial(x_A,\theta,u)}=1\,.
\label{Ber}
\ee
Using the coordinate transformations (\ref{dxA}) and (\ref{du}),
it is straightforward to find the superconformal variations  of harmonic derivatives:
\be
\begin{array}{ll}
\delta_{\rm sc} D^1_2=-\lambda^1_2 S_1\,, & \delta_{\rm sc} D^2_1=(\lambda^1_1-\lambda^2_2)D^2_1\,,\\
\delta_{\rm sc} D^2_3=-\lambda^2_3 S_2\,, & \delta_{\rm sc} D^3_2=(\lambda^2_2-\lambda^3_3)D^3_2\,,\\
\delta_{\rm sc} D^1_3=\lambda^1_2D^2_3-\lambda^2_3D^1_2-\lambda^1_3(S_1+S_2)\,,\quad &
\delta_{\rm sc} D^3_1=(\lambda^1_1-\lambda^3_3)D^3_1+\lambda^2_1D^3_2-\lambda^3_2D^2_1\,,\\
\delta_{\rm sc} D^1_1=\delta_{\rm sc} D^2_2=\delta_{\rm sc} D^3_3=0\,,&
\delta_{\rm sc} S_1=\delta_{\rm sc} S_2=0\,.
\end{array}
\label{dD}
\ee
Recall that the gauge covariant harmonic derivatives involve the gauge superfield prepotentials
\be
\nabla^I_J=D^I_J+iV^I_J\,.
\label{cov-gauge}
\ee
Requiring the lengthened derivatives (\ref{cov-gauge}) to be superconformally covariant, with taking into account the transformations (\ref{dD}),
implies the following transformation laws for the gauge prepotentials:
\be
\begin{array}{ll}
\delta_{\rm sc} V^1_2=0\,, & \delta_{\rm sc} V^2_1=(\lambda^1_1-\lambda^2_2)V^2_1\,,\\
\delta_{\rm sc} V^2_3=0\,, & \delta_{\rm sc} V^3_2=(\lambda^2_2-\lambda^3_3)V^3_2\,,\\
\delta_{\rm sc} V^1_3=\lambda^1_2 V^2_3-\lambda^2_3 V^1_2\,, \quad&
 \delta_{\rm sc} V^3_1=(\lambda^1_1-\lambda^3_3)V^3_1+
  \lambda^2_1V^3_2-\lambda^3_2V^2_1\,.
\end{array}
\label{dV}
\ee
Note that the superconformal variations of the analytic gauge
superfields $V^1_2$, $V^2_3$ and $V^1_3$ were earlier given
in \cite{GIO-N3,hss-book}, while the transformations of the
non-analytic gauge superfields $V^2_1$, $V^3_2$ and $V^3_1$ were not presented before.

Using (\ref{dxA}) and (\ref{du}) it is also easy to find the
superconformal transformations of the covariant spinor derivatives
(\ref{anal-deriv}),
\bea
\delta_{\rm sc} D^1_\alpha&=&(-a/2-ib-\lambda^1_1)D^1_\alpha+B_\alpha^\beta D^1_\beta\,,
\nn\\
\delta_{\rm sc}\bar D_{3\dot\alpha}&=&(-a/2+ib+\lambda^3_3)\bar D_{3\dot\alpha}+
 \bar B_{\dot\alpha}^{\dot\beta}\bar D_{3\dot\beta}\,,
\label{delta-D}
\eea
where $\lambda^1_1$ and $\lambda^3_3$ were defined in
(\ref{lambda}) and
\bea
B_\alpha^\beta&=&-k_{\alpha\dot\beta}(x_{A+}^{\beta\dot\beta}+
    4i\theta_1^\beta\bar\theta^{1\dot\beta})-4i\theta_I^\beta
    u^I_j\eta^j_\alpha\,,\nn\\
\bar B_{\dot\alpha}^{\dot\beta}&=&-k_{\beta\dot\alpha}
   (x_{A-}^{\beta\dot\beta}-4i\theta_3^\beta\bar\theta^{3\dot\beta})-
   4i\bar\theta^{I\dot\beta}\bar u_I^j\bar\eta_{\dot\alpha j}\,.
\eea
It is worth pointing out that the spinor derivatives $ D^1_\alpha $ and $\bar D_{3\dot\alpha}$ are not mixed under the superconformal
transformations.

Finally, using the variations of the gauge prepotentials
(\ref{dV}) and derivatives (\ref{delta-D}), we can find the
superconformal transformations of the superfield strengths
(\ref{W-anal}),
\be
\delta_{\rm sc} W_{23}=A W_{23}\,,\qquad
\delta_{\rm sc} \bar W^{12}=\bar A \bar W^{12}\,,
\label{dW}
\ee
where
\be
A=-a+2ib+\lambda^2_2+\lambda^3_3+\bar
B^{\dot\alpha}_{\dot\alpha}\,,\qquad
\bar A=-a-2ib-\lambda^1_1-\lambda^2_2+B^\alpha_\alpha\,.
\label{A}
\ee
One can check that the superfields $A$ and $\bar A$ are analytic,
\be
D^1_\alpha (A,\;\bar A)=\bar D_{3\dot\alpha}(A,\;\bar A)=0\,.
\ee
Hence, the transformations (\ref{dW}) preserve analyticity.

\section{Superconformal effective action}
\subsection{Non-superconformal $F^4$ term}
The $\cN=3$ supersymmetric completion of the fourth-order term
in the Born-Infeld action was constructed in \cite{IZ},
\be
S_4=\frac1{32}\int  d\zeta(^{33}_{11})du\,
\frac{(\bar W^{12} W_{23})^2}{(\bar\Lambda\Lambda)^2}\,.
\label{S4}
\ee
Here $\Lambda$ is a coupling constant of dimension one in mass units, which is
introduced to ensure the correct dimension of the integrand.
The analytic measure is defined as follows \cite{IZ,BISZ},
\be
d\zeta(^{33}_{11})=\frac1{16^2}d^4x_A(D^3)^2(D^2)^2(\bar D_1)^2(\bar
D_2)^2\,.
\label{measure}
\ee
The analytic measure is dimensionless,
$[d\zeta(^{33}_{11})du]=0\,$, and $[\bar W^{12}]=[W_{23}]=1\,$.
With this normalization of the analytic measure, it is
straightforward to check that, along with other component terms, the
action (\ref{S4}) yields the standard $F^4$ term,
\be
S_4=\frac12\int d^4x\,\frac{F^2\bar
F^2}{(\bar\Lambda\Lambda)^2}+\ldots\,.
\label{F4}
\ee

Consider now the superconformal variation of the action (\ref{S4}),
\be
\delta_{\rm sc} S_{4}=\frac1{16}\int  d\zeta(^{33}_{11})du(A+\bar A)
\frac{(\bar W^{12} W_{23})^2}{(\bar\Lambda\Lambda)^2}\,,
\label{dS4}
\ee
where we have used the variations of the superfield strengths
(\ref{dW}) and the invariance of the analytic measure (\ref{Ber}).
Here $A$ and $\bar A$ are superfields (\ref{A}) collecting
the constant parameters of the superconformal transformations
(\ref{dxA}) and (\ref{du}). The variation (\ref{dS4}) is non-zero, hence the action (\ref{S4})
is not superconformal.

\subsection{Scale and $\gamma_5$ invariant $F^4/X^4$ term}
Our aim here is to find a superconformal generalization of the action (\ref{S4}).
In what follows we will denote this superconformal action by $\Gamma$ (to stress that
it is a part of the $\cN=3$ SYM low-energy effective action). The action $\Gamma$
should meet the following criteria:
\begin{enumerate}
\item It should be a local functional defined on the analytic superspace and constructed out of the superfield strengths
$\bar W^{12}$ and $W_{23}$ without derivatives on them,
\be
\Gamma=\int d\zeta(^{33}_{11}) du\,{\cal H}^{11}_{33}(\bar W^{12},W_{23})\,.
\label{3.5}
\ee
The analytic Lagrangian density ${\cal H}^{11}_{33}$ is an arbitrary function of its
arguments, such that its external harmonic U(1) charges cancel those of the analytic integration measure.
This is the most general form of the superspace action yielding terms with four-derivatives in components,  since
the analytic measure (\ref{measure}) contains just eight spinor derivatives which can produce four space-time ones
on the component fields.
\item
The action $\Gamma$ should be invariant under the superconformal
transformations (\ref{dW}),
\be
\delta_{\rm sc} \Gamma=0\,.
\ee
As a weaker requirement, in this subsection we will employ only the scale- and
$\gamma_5$-transformations out of the full SU(2,2$|$3)
superconformal group. We will show that this is sufficient to uniquely specify the structure of the action. The check of the full
superconformal symmetry will be performed in the next subsection.
\item
In the component-field expansion the action $\Gamma$ should
reproduce the scale- and SU(3)-invariant
$F^4/X^4$ term (\ref{F4}),
\be
\int d^4x\frac{F^2\bar F^2}{(\varphi^i\bar\varphi_i)^2}\,.
\ee
\item
We are interested in the low-energy effective action for massless
fields, with massive ones being integrated out. The massive fields appear
in the Coulomb branch when the gauge symmetry is broken down
spontaneously. For instance, the SU(2) gauge symmetry is broken down to U(1)
when the scalar field corresponding to the Cartan subalgebra of
su(2) acquire non-trivial vevs,
\be
c^i=\langle \varphi^i \rangle\ne0\,,\qquad
\bar c_i=\langle \bar \varphi_i \rangle \ne0\,.
\label{c}
\ee
However, the effective action should be independent of any particular choice
of these constants,
\be
\Gamma({c'}^i,\bar c'_j)=\Gamma(c^i,\bar c_j)\,, \qquad c^i\bar c_i \ne0\,,
\ee
because such a dependence would break superconformal invariance of the action.
\item
Finally, we simplify the problem by considering only that part of
the action (\ref{3.5}) which does not vanish on the mass shell, i.e.,
we will assume that the superfield strengths obey the constraints
(\ref{anal})--(\ref{linearity}). We will neglect all terms in
the action $\Gamma$ which vanish when these constraints are imposed.
As a consequence, one is free to add to $\Gamma$ or to subtract from it
the following expressions which vanish on the mass shell,
\bea
\int d\zeta(^{33}_{11})\, \bar W^{12}{\cal F}(W_{23})&\propto&
\int d^4x (D^3)^2 (D^2)^2 (\bar D_1)^2[{\cal F}(W_{23})  (\bar D_2)^2 \bar
W^{12}] \simeq 0\,,\nn\\
\int d\zeta(^{33}_{11})\, W_{23}{\cal F}( \bar W^{12})&\propto&
\int d^4x (D^3)^2 (\bar D_2)^2 (\bar D_1)^2[{\cal F}( \bar W^{12})
(D^2)^2 W_{23}] \simeq 0\,.\nn\\
\label{prop}
\eea
Here ${\cal F}(W)$ is an arbitrary function of its argument.
We will frequently employ this property while deriving the action.
\end{enumerate}

Now we shall turn to constructing the action $\Gamma$ which obeys the requirements and properties listed above.

As the first step, we introduce the shifted scalar fields, $\phi^i$ and
$\bar\phi_i$,
\be
\varphi^i=c^i+\phi^i\,,\quad
\bar\varphi_i=\bar c_i+\bar \phi_i\,,\qquad
\langle \phi^i \rangle=\langle \bar\phi_i \rangle=0\,.
\label{vev-shift}
\ee
Next, we define the harmonic projections of these vev
constants
\be
c^1=u^1_i c^i\,,\quad c^2=u^2_i c^i\,\quad
c^3=u^3_i c^i\,,\qquad
\bar c_1 =\bar u_1^i \bar c_i\,,\quad
\bar c_2 =\bar u_2^i \bar c_i\,,\quad
\bar c_3 = \bar u_3^i \bar c_i\,.
\label{cccc}
\ee
Using these objects, we introduce the shifted superfield
strengths, $\bar\omega^{12}$ and $\omega_{23}$,
\be
\bar W^{12}=\bar c_3+\bar\omega^{12}\,,\qquad
W_{23}=c^1+\omega_{23}\,.
\label{shift}
\ee
Under the scale and $\gamma_5$ transformations these shifted
superfields transform inhomogeneously,
\be
\delta_{\rm sc} \bar\omega^{12}=\bar A\bar c_3 +\bar A\bar\omega^{12}\,,\qquad
\delta_{\rm sc} \omega_{23}=Ac^1+A\omega_{23}\,,
\label{scale}
\ee
where $A=-a+2ib$. The case of generic $A$ and $\bar A$ defined in (\ref{A}) will be considered in the next
subsection.

We point out that on shell, when the relations (\ref{prop}) are valid, the non-superconformal action (\ref{S4}) can be
rewritten in terms of $\bar \omega^{12}$ and $\omega_{23}$ as
\be
S_4=\frac1{32}\int  d\zeta(^{33}_{11})du\,
\frac{(\bar \omega^{12} \omega_{23})^2}{(c^i\bar c_i)^2}\,.
\label{S4_}
\ee
Here we substituted $(c^i\bar c_i)^2$ in the denominator instead of
$(\bar\Lambda\Lambda)^2$, because no other dimensionful constants besides the vevs $c^i$ can be
present in the superconformal case.

We search for a superconformal generalization of the action (\ref{S4_}) in the form
\be
\Gamma=\frac\alpha{8}\int  d\zeta(^{33}_{11})du\frac{(\bar \omega^{12}
\omega_{23})^2}{(c^i\bar c_i)^2}
H\left(\frac{\bar \omega^{12}c^3}{c^i \bar c_i},
\frac{\omega_{23}\bar c_1}{c^i\bar c_i}\right),
\label{G}
\ee
where $H(x,y)$ is some function to be determined and $\alpha$ is a
dimensionless coupling constat.
The arguments $\frac{\bar \omega^{12}c^3}{c^i \bar
c_i}$ and $\frac{\omega_{23}\bar c_1}{c^i\bar c_i}$ of the function $H$ are chargeless and dimensionless.
We assume that the function $H$ has a regular power expansion with respect to its arguments,
\be
H(x,y)=\sum_{m,n=0}^\infty
\alpha_{m,n} x^m y^n\,,
\label{H}
\ee
with undefined coefficients $\alpha_{m,n}$.
The reality of the action (\ref{G}) under complex conjugation
implies the symmetry of this function, $H(x,y)=H(y,x)\,$, whence $\alpha_{m,n}=\alpha_{n,m}\,$.

Reordering the summation in (\ref{H}), it is convenient to
represent (\ref{G}) as
\be
\Gamma=\sum_{n=0}^\infty\Gamma_n\,,\qquad
\Gamma_n=\frac\alpha{8}\int d\zeta(^{33}_{11})du
\frac{(\bar \omega^{12}\omega_{23})^2}{(c^i\bar c_i)^2}
\sum_{i=0}^{n}
\alpha_{i,n-i}
\left(\frac{\bar\omega^{12}c^3}{c^i\bar c_i}\right)^i
\left(\frac{\omega_{23}\bar c_1}{c^i\bar c_i}\right)^{n-i}.
\label{Gn}
\ee
The invariance of the action (\ref{Gn}) under the transformations
(\ref{scale}) can be ensured order by order, i.e., the
non-vanishing terms from $\delta_{\rm sc} \Gamma_{n}$ are required to be cancelled by similar terms from $\delta_{\rm sc} \Gamma_{n+1}$,
and so forth.  This recurrence procedure imposes severe restrictions on the coefficients
$\alpha_{m,n}$. The technical details of this procedure are given in the Appendix A, with the following result:
\be
\alpha_{m,n}=(-1)^{m+n}\frac{(m+n+2)!}{(n+2)n!(m+2)m!}\,.
\label{alpha}
\ee
With these coefficients, the series (\ref{H}) can be summed up as follows,
\be
H(x,y)=
  \frac{\ln (1 + x + y)}{x^2y^2}+
\frac{1}
   {xy( 1 + x +
       y ) } -
  \frac{\ln (1 + x)}{x^2 y^2} -
  \frac{\ln (1 + y)}{x^2y^2} \,.
\label{HH}
\ee
We point out that this function is regular at the origin,
\be
\lim_{x,y\to0}H(x,y)=\frac12\,.
\ee
Hence, the action (\ref{G}) with this function is well-defined and
the harmonic integral does not encounter any singularities.

The contributions from the last two terms in (\ref{HH}) to the
action (\ref{G}) vanish on shell due to the properties
(\ref{prop})\footnote{The properties (\ref{prop}) are valid essentially on shell.
Therefore the last two terms in (\ref{HH}) can be neglected only
on shell although they can be important for the off-shell
completion of the action.}. Therefore, the on-shell effective
action can be rewritten in the following explicit form
\be
\Gamma=\frac\alpha{8}\int d\zeta(^{33}_{11})du\left[\frac{(c^i\bar c_i)^2}{c^3c^3 \bar c_1 \bar c_1}
\ln\left(1+\frac{\bar\omega^{12}c^3}{c^i\bar c_i}
+\frac{\omega_{23}\bar c_1}{c^i\bar c_i}\right)
+\frac{(c^i\bar c_i)\bar\omega^{12}\omega_{23}}{c^3\bar c_1(c^i\bar c_i
+\bar\omega^{12}c^3+\omega_{23}\bar c_1)}
\right].
\label{Gconf}
\ee
Although the charged objects $c^3$ and $\bar c_1$ appear in the
denominators, they do not lead to the divergent harmonic
integrals. It can be explicitly checked  that upon passing
to the component form of the action (\ref{Gconf}), all dangerous
terms with divergent harmonic integrals vanish after performing the integration
over the Grassmann variables. Some component terms of this action
will be studied in the next Section.

\subsection{Complete $\cN=3$ superconformal symmetry}

Now we consider the transformations (\ref{dW})
which include all parameters of the superconformal
transformations. The corresponding variations (\ref{scale}) of the
shifted superfield strengths $\bar \omega^{12}$ and $\omega_{23}$
read
\bea
\delta_{\rm sc} \bar \omega^{12}&=&A\bar\omega^{12}+A\bar c_3+\lambda^2_3
\bar c_2+\lambda^1_3 \bar c_1\,,\nn\\
\delta_{\rm sc} \omega_{23}&=&\bar A \omega_{23}+\bar A c^1
-\lambda^1_2 c^2-\lambda^1_3 c^3\,,
\eea
where $A$ and $\bar A$ are given in (\ref{A}) and $\lambda^I_J$
are defined in (\ref{lambda}). Under these transformations the
action (\ref{G}) varies as
\bea
\delta_{\rm sc} \Gamma&=&\frac\alpha{8}\int d\zeta(^{33}_{11})du\, (\bar\omega^{12}\omega_{23})^2
[\frac{2}x H(x,y)+H'_x(x,y)][Ax+Ac^3\bar c_3+\lambda^2_3 c^3\bar c_2
+\lambda^1_3 c^3 \bar c_1]
\nn\\
&&+\,\frac\alpha{8}\int d\zeta(^{33}_{11})du\, (\bar\omega^{12}\omega_{23})^2
[\frac{2}y H(x,y)+H'_y(x,y)][\bar Ay+\bar Ac^1\bar c_1-\lambda^1_2 c^2\bar
c_1 -\lambda^1_3 c^3 \bar c_1]\,.\nn\\
\label{var1}
\eea
For simplicity we set here $c^i\bar c_i=1\,$, so $x=\bar\omega^{12}c^3$, $y=\omega_{23}\bar c_1$. The
first and second lines in (\ref{var1}) are complex-conjugated to each other.

Given the explicit form (\ref{HH}) of the function $H(x,y)$, it is
easy to check that it solves the following differential equations
\bea
\frac2xH(x,y)+H'_x(x,y)&=&\frac{1}{x(1+x)(1+x+y)^2}\,,\nn\\
\frac2yH(x,y)+H'_y(x,y)&=&\frac{1}{y(1+y)(1+x+y)^2}\,.
\label{relH}
\eea
Taking them into account, we are going to show that the integrand in (\ref{var1}) is a total harmonic derivative,
so the variation (\ref{var1}) vanishes.

To this end, we introduce the auxiliary functions $f(x,y)$ and $\tilde{f}(x,y)$:
\bea
f(x,y)&=&\frac{1}{y(y+1)(x+y+1)}+\frac{\ln(1+x+y)}{xy^2}
-\frac{\ln(1+x)}{xy^2}-\frac{\ln(1+y)}{xy^2}\,,\\
\tilde f(x,y)&=&f(y,x)=
\frac{1}{x(x+1)(x+y+1)}+\frac{\ln(1+x+y)}{yx^2}
-\frac{\ln(1+y)}{yx^2}-\frac{\ln(1+x)}{yx^2}\,.\nn
\label{f}
\eea
They possess the following properties
\bea
xf'_x+f&=&-\frac1{(1+x)(1+x+y)^2}=-(xH'_x+2H)\,,
\label{prop0.1}\\
xf'_x+yf'_y+3f&=&\frac1{x(1+x)(1+x+y)^2}-\frac1{x(1+y)^2}
=(H'_x+\frac 2xH)+\ldots\,,\label{prop1.1}\\
y\tilde f'_y+\tilde f&=&-\frac1{(1+y)(1+x+y)^2}=-(yH'_y+2H)\,,\\
y\tilde f'_y+x\tilde f'_x +3\tilde f&=&
\frac1{y(1+y)(1+x+y)^2}-\frac1{y(1+x)^2}=(H'_y+\frac 2x
H)+\ldots\,.
\label{prop1}
\eea
Here dots stand for the terms integrals of which over the
analytic superspace with the weight
$(\bar\omega^{12}\omega_{23})^2$ are on-shell vanishing due to the relations (\ref{prop}).
Up to these terms, the
equations (\ref{prop0.1})--(\ref{prop1}) allow one to deduce the following relations
\bea
-D^2_3 (f(x,y)c^3 \bar c_2 A)- D^1_3(f(x,y)c^3 \bar c_1 A)
&=&(H'_x+\frac 2x H)(Ax+Ac^3\bar c_3)
\nn\\&&
-\,f(x,y)c^3 \bar c_2
\lambda^2_3
-f(x,y)c^3\bar c_1 \lambda^1_3\,,\nn\\
D^1_2(\tilde f(x,y)c^2\bar c_1\bar A)
+D^1_3(\tilde f(x,y)c^3\bar c_1\bar A)
&=&(H'_y+\frac2yH)(\bar A y+\bar A c^1\bar c_1)
\nn\\&&
+\,\tilde f(x,y)c^2\bar c_1\lambda^1_2
+\tilde f(x,y)c^3\bar c_1\lambda^1_3\,.
\label{rel1}
\eea
Here we made use of the following simple identities
\be
\lambda^1_2=D^1_2\bar A\,,\quad
\lambda^2_3=D^2_3A\,,\quad
\lambda^1_3=D^1_3A=D^1_3\bar A\,,
\ee
as well as  of the convention $c^i\bar c_i=1\,$.

Next, we introduce the functions
\bea
g(x,y)&=&\frac1{y(1+y)^2(1+x+y)}-\frac1{y(x+1)}\,,\\
\tilde g(x,y)&=&g(y,x)=\frac1{x(1+x)^2(1+x+y)}-\frac1{x(y+1)}\,,
\label{g}
\eea
with the properties
\bea
xg'_x+g&=&\frac1{y(1+y)(1+x+y)^2}-\frac1{y(1+x)^2}
=H'_y+\frac2yH+\ldots\,,\\
y\tilde g'_y+\tilde
g&=&\frac1{x(1+x)(1+x+y)^2}-\frac1{x(1+y)^2}
=H'_x+\frac 2xH+\ldots\,,\\
g(x,y)-\tilde g(x,y)&=&(H'_x+\frac 2x H)-(H'_y+\frac 2y H)\,.
\eea
Here, as in (\ref{prop1.1}) and in (\ref{prop1}), the dots stand for the terms vanishing on shell after integration over the
analytic superspace with the weight $(\bar
\omega^{12}\omega_{23})^2$. Up to these terms, we obtain the
following relation
\bea
-D^1_2(\lambda^2_3\tilde g(x,y)c^3\bar c_1)-D^2_3(\lambda^1_2 g(x,y)c^3\bar c_1)
&=&(H'_x+\frac2xH)\lambda^2_3 c^3\bar c_2
-(H'_y+\frac 2yH)\lambda^1_2c^2\bar c_1
\nn\\&&+\,[(H'_x+\frac2xH)-(H'_y+\frac 2yH)]
\lambda^1_3 c^3\bar c_1\,.
\nn\\
\label{rel2}
\eea

Finally, introduce the functions
\bea
h(x,y)&=&-\frac1{(1+x)y}+\frac{\ln(1+x)}{xy^2}+\frac{\ln(1+y)}{xy^2}
-\frac{\ln(1+x+y)}{xy^2}\,,
\\
\tilde h(x,y)&=&h(y,x)=
-\frac1{(1+y)x}+\frac{\ln(1+y)}{yx^2}+\frac{\ln(1+x)}{yx^2}
-\frac{\ln(1+x+y)}{yx^2}\,,
\label{h}
\eea
with the properties
\bea
&h(x,y)+yh'_y(x,y)=f(x,y)\,,\qquad
\tilde h(x,y)+x\tilde h'_x(x,y)=\tilde f(x,y)\,,&\\
&h-\tilde h=\tilde f-f\,.&
\eea
These properties allow one to derive the following relation
\be
-D^1_2(\lambda^2_3 h(x,y)c^3\bar c_1)-D^2_3(\lambda^1_2 \tilde h(x,y)c^3\bar c_1)
=f\lambda^2_3 c^3\bar c_2
-\tilde f\lambda^1_2c^2\bar c_1
+(f-\tilde f)\lambda^1_3 c^3\bar c_1\,.
\label{rel3}
\ee

Now we put together the relations (\ref{rel1}), (\ref{rel2}) and
(\ref{rel3}) and observe that the variation (\ref{var1}) can
be represented as a linear combination of harmonic
derivatives acting on the quantities which are expressed through the
functions (\ref{f}), (\ref{g}) and (\ref{h}),
\bea
\delta_{\rm sc} \Gamma&=&\frac\alpha{8}\int d\zeta(^{33}_{11})du\, (\bar\omega^{12}\omega_{23})^2
\bigg\{
D^1_2(\tilde f c^2 \bar c_1 \bar A)
- D^2_3( f c^3 \bar c_2 A)
+D^1_3(\tilde f c^3 \bar c_1\bar A- f c^3 \bar c_1 A)
\nn\\&&
-\,D^1_2[(\tilde g+h)\lambda^2_3 c^3\bar c_1]-D^2_3[(g+\tilde h)\lambda^1_2 c^3\bar c_1]
\bigg\}.\label{344}
\eea
The variation (\ref{344}) vanishes as an integral of total
harmonic derivative. This proves the invariance of the action
(\ref{Gconf}) under the full SU(2,2$|$3) superconformal group\footnote{Note that (\ref{Gconf}) is SU(2,2$|$3) invariant for any $c^i\neq 0\,$,
without any restriction on the norm $c^i\bar c_i$ which was put equal to $1$ in the above consideration merely for convenience.}.

\subsection{Independence of the choice of vacua}
By construction, the effective action (\ref{G}) with the function
(\ref{HH}) is meaningful only on the Coulomb branch of the $\cN=3$
SYM theory. This is manifested in the explicit presence of non-zero
vev constants $c^i$ and $\bar c_i$ in the Lagrangian in (\ref{G}). However, the
action itself should be independent of any particular choice of these
constants, except for the point $c^i=0$ at which the effective action
is singular.

Let us rewrite (\ref{G}) in terms of the original
(non-shifted) superfield strengths $\bar W^{12}$ and $W_{23}$
\be
\Gamma[\bar W^{12},W_{23};c^i,\bar c_i]=\frac\alpha{8}\int  d\zeta(^{33}_{11})du\frac{(\bar W^{12}-\bar c_3)^2
(W_{23}-c^1)^2}{(c^i\bar c_i)^2}
H\left(c^3\frac{\bar W^{12}-\bar c_3}{c^i \bar c_i},
\bar c_1\frac{W_{23}-c^1}{c^i\bar c_i}\right).
\label{GG}
\ee
In the previous subsection we proved that this action is invariant
under the full SU(2,2$|$3) superconformal group. Taking into account that the analytic integration measure is SU(2,2$|$3) invariant by itself,
the property of superconformal invariance of the action can be written in the finite form as
\be
\Gamma[\bar W^{12},W_{23};c^i,\bar c_i] = \Gamma'[\bar W^{12}{}',W_{23}{}';c^i,\bar c_i] = \Gamma[\bar W^{12}{}',W_{23}{}';c^i,\bar c_i]\,.
\ee
In particular,
consider scale and $\gamma_5$ transformations of the superfield
strength in the finite form,
\be
\bar W^{12}\to e^{\bar A} \bar W^{12}\,,\qquad
W_{23}\to e^A W_{23}\,,
\label{scale-finite}
\ee
where $A=-a+2ib$. The transformation of the action (\ref{GG})
under (\ref{scale-finite}) can be represented as
\bea
\Gamma[\bar W^{12},W_{23};c^i,\bar c_i] &=& \Gamma[e^{\bar A} \bar W^{12},e^A W_{23};c^i,\bar c_i] \nonumber \\
&=&\frac\alpha{8}
\int  d\zeta(^{33}_{11})du\frac{(\bar W^{12}-e^{-\bar A}\bar c_3)^2
(W_{23}-e^{-A}c^1)^2}{(e^{-A-\bar A}c^i\bar c_i)^2}
\nn\\&&\times
\,H\left(e^{-A}c^3\frac{\bar W^{12}-e^{-\bar A}\bar c_3}{e^{-A-\bar A}c^i \bar c_i},
e^{-\bar A}\bar c_1\frac{W_{23}-e^{-A}c^1}{e^{-A-\bar A}c^i\bar c_i}\right).
\eea
So, all $A$-dependence is absorbed into the vev constants,
$c^i\to e^{-A}c^i$, $\bar c_i\to e^{-\bar A}\bar c_i$. Hence, the
superconformal invariance of the action (\ref{GG}) implies its
independence of complex rescalings of the vev constants,
\be
\Gamma[\bar W^{12},W_{23};c^i,\bar c_i]
=\Gamma[e^{\bar A} \bar W^{12},e^A W_{23};c^i,\bar c_i]
=\Gamma[ \bar W^{12},W_{23};e^{-\bar A}c^i,e^{-A}\bar c_i]\,.
\ee

In a similar way, one can prove that the action
(\ref{GG}) is independent of the parameters of finite SU(3) rotations of the vev constants,
\be
\Gamma[\bar W^{12},W_{23};c^i,\bar c_i]
=\Gamma[ \bar W^{12},W_{23};\Lambda^i_j c^j,\bar \Lambda_i^j\bar
c_j]\,,
\ee
where $\Lambda^i_j$ are SU(3) matrices. As a result, the action
(\ref{GG}) is independent of any particular choice of the vacuum
$c^i$, $c^i\ne0\,$.

Perhaps, it make sense to give a more detailed proof of the latter statement. It goes as follows. Let us assume, without loss of generality,
that $c^3\neq 0\,$. Then, using the coset SU(3)/[U(1)$\times $SU(2)] transformations with a constant SU(2) doublet as parameters, one can
cast $c^i$ in the form
$c^i = (0,0,c^3)$. The constant $c^3$ can be made real by making use of the residual U(1) transformation
(a combination of the $\gamma_5$ transformations
and those of U(1) from the denominator of SU(3)/[U(1)$\times $SU(2)]). Finally, it can be rescaled to any non-zero value,
keeping in mind the independence of the action of the rescalings of the vev constants.

\section{Component structure}
\subsection{$F^4/X^4$ term}
To derive this term from the effective action (\ref{G}), it is
sufficient to consider only constant Maxwell and scalar fields,
omitting all other components in (\ref{W-comp}),
\be
\hat{\bar\omega}^{12}=u^1_i \phi^i+4i\theta_2^\alpha\theta_3^\beta
F_{\alpha\beta}\,,\qquad
\hat\omega_{23}=\bar u_3^i
\bar\phi_i+4i\bar\theta^{1\dot\alpha}\bar\theta^{2\dot\beta}\bar
F_{\dot\alpha\dot\beta}\,.
\ee
Substituting these superfields into (\ref{G}), we integrate over the
Grassmann variables to obtain
\be
\Gamma_{F^4/X^4}=\frac\alpha{2}\int d^4x du\, F^2\bar F^2
\sum_{m,n=0}^\infty \frac{(m+1)(n+1)(m+n+2)!(-1)^{m+n}}{m!n!}
(\bar \phi_3 c^3)^m (\phi^1 \bar c_1)^n\,.
\label{G-comp}
\ee
Here we used the series expansion (\ref{H}) for the function
$H$ with the coefficients given by (\ref{alpha}). In this
subsection we assume $c^i\bar c_i=1$ for simplicity
and use the notation
$F^2=F^{\alpha\beta}F_{\alpha\beta}$\,,
$\bar F^2=\bar F^{\dot\alpha\dot\beta}\bar
F_{\dot\alpha\dot\beta}$\,.

It is convenient to represent (\ref{G-comp}) as a sum of two
terms,
\be
\Gamma_{F^4/X^4}=\frac\alpha2\int d^4x\, F^2\bar F^2(T_1+T_2)\,,
\ee
where
\bea
T_1&=&\int du\,
\sum_{n=0}^\infty \sum_{m=0}^n \frac{(m+1)(n+1)(m+n+2)!(-1)^{m+n}}{m!n!}
(\phi^1 \bar c_1)^n (\bar \phi_3 c^3)^m \,,\nn\\
T_2&=&\int du\,
\sum_{n=0}^\infty \sum_{m=n+1}^\infty \frac{(m+1)(n+1)(m+n+2)!(-1)^{m+n}}{m!n!}
(\phi^1 \bar c_1)^n(\bar \phi_3 c^3)^m \,.
\eea
The reason for this separation is that $m\leq n$ in $T_1$ while $m>n$
in $T_2$. Therefore, for each of these terms we can apply the
equation (\ref{hint}) for the harmonic integrals,
\bea
T_1&=&2\sum_{n=0}^\infty \sum_{m=0}^n \sum_{l=0}^m
\frac{(m+n-l+1)!(-1)^{m+n+l}}{l!(n-l)!(m-l)!}
(\phi^i\bar\phi_i)^l (\phi^i\bar
c_i)^{n-l}(c^i\bar\phi_i)^{m-l}\,,\nn\\
T_2&=&2\sum_{n=0}^\infty \sum_{m=n+1}^\infty \sum_{l=0}^n
\frac{(m+n-l+1)!(-1)^{m+n+l}}{l!(n-l)!(m-l)!}
(\phi^i\bar\phi_i)^l (\phi^i\bar
c_i)^{n-l}(c^i\bar\phi_i)^{m-l}\,.
\eea
Changing the order of summation, these
terms can be rewritten as
\bea
T_1&=&2\sum_{l,m=0}^\infty \sum_{n=m}^\infty
\frac{(n+m+l+1)!(-1)^{m+n+l}}{l!m!n!}(\phi^i\bar\phi_i)^l
(\phi^i\bar\phi_i)^n (c^i\bar\phi_i)^m\,,\nn\\
T_2&=&2\sum_{l,m=0}^\infty \sum_{n=0}^{m-1}
\frac{(n+m+l+1)!(-1)^{m+n+l}}{l!m!n!}(\phi^i\bar\phi_i)^l
(\phi^i\bar\phi_i)^n (c^i\bar\phi_i)^m\,.
\eea
Putting these two expressions together, we find
\bea
T_1+T_2&=&2\sum_{m,n,k=0}^\infty
\frac{(-1)^{m+n+k}(m+n+k+1)!}{m!n!k!}
(c^i\bar\phi_i)^{m}(\bar c_i\phi^i)^n(\bar\phi_i\phi^i)^{k}
\nn\\&
=&\frac2{(1+c^i\bar\phi_i+\bar c_i \phi^i+\phi^i\bar\phi_i)^2}
=\frac2{(\varphi^i\bar\varphi_i)^2}\,.
\eea
As a result, the $F^4/X^4$ term in the effective action reads
\be
\Gamma_{F^4/X^4}=\alpha\int d^4x\frac{F^2\bar
F^2}{(\varphi^i\bar\varphi_i)^2}\,. \label{F4X}
\ee
This expression is explicitly scale and U(3) invariant, as is
expected.

It is a highly non-trivial and remarkable phenomenon that the vev constants $c^i$ and the
shifted scalars $\phi^i$ have combined into the initial scalar
fields $\varphi^i$, (\ref{vev-shift}), after doing the Grassmann and harmonic integrals which
is a rather involved procedure in its own.
This confirms the independence of the action (\ref{G}) of any particular choice of
the vacua, the fact that was proved in the previous section.

Note that (\ref{F4X}) also respects hidden SO(6)$\simeq$ SU(4) invariance, with the SU(4)$/$U(3) transformations acting as
\be
\delta \varphi^i = \varepsilon^{ikl}\lambda_k\bar\varphi_l\,, \quad
\delta \bar\varphi_i = \varepsilon_{ikl}\bar\lambda^k\varphi^l\,,
\ee
where $\lambda_i$ comprise 6 corresponding group parameters. This is an indication that the superfield effective action (\ref{G}), besides
the superconformal SU(2,2$|$3) symmetry, enjoys as well an on-shell SU(4) symmetry, and hence,  the
superconformal SU(2,2$|$4) symmetry as a closure of the two former ones. It would be interesting to explicitly find the realization
of this SU(4) on the analytic superfield strengths.

\subsection{$F^6/X^8$ term}
It is known that the non-conformal action (\ref{S4}) produces
not only the $F^4$ term but also the $F^6$ term in
the Born-Infeld action \cite{IZ}. The $F^6$ term appears
essentially on shell, when some of the auxiliary fields are
eliminated by their effective equations of motion. In \cite{BISZ}
it was conjectured that this procedure should work in a similar way in the
superconformal case, with the $F^6/X^8$ term as the outcome. Here
we show that this is indeed the case and present details of the relevant derivation.

As shown in \cite{IZ}, the Maxwell field strength $F_{mn}=\partial_m A_n-\partial_n A_m$
is accompanied by the antisymmetric tensor auxiliary field $H_{mn}=-H_{nm}$.
When the on-shell constraints are relaxed, these fields appear in the superfield
strengths in the combination $V_{mn}=\frac14(F_{mn}+H_{mn})$,
\be
\bar\omega^{12}=u^1_i \phi^i+4i\theta_2^\alpha\theta_3^\beta
V_{\alpha\beta}+\ldots\,,\qquad
\omega_{23}=\bar u_3^i
\bar\phi_i+4i\bar\theta^{1\dot\alpha}\bar\theta^{2\dot\beta}\bar
V_{\dot\alpha\dot\beta}+\ldots\,.
\label{W+aux}
\ee
Here $V_{\alpha\beta}$ and $\bar V_{\dot\alpha\dot\beta}$ are
spinorial components of the antisymmetric tensor $V_{mn}$ and dots
stand for the other field components which are irrelevant for
our consideration.
The part of the free classical action $S_2$ which involves
these fields reads \cite{IZ}
\be
S_2=\int d^4x[V^2+\bar V^2-2(VF+\bar V \bar F)+\frac12(F^2+\bar
F^2)]\,.
\label{S2}
\ee
One can recover the standard Maxwell action for
$F_{mn}$ upon eliminating the auxiliary fields from $S_2$.
However, our purpose is to eliminate them from the {\it effective}
equations of motion, when the action (\ref{G}) is added to the
classical free SYM action. The superfield effective equations
of motion are derived in Appendix D. Here we need only some SU(3) singlet
sub-sector of the component expansion of these equations, so
it is simpler to derive it independently.

As in the previous subsection, we substitute the
superfields (\ref{W+aux}) into (\ref{G}) and find
\be
\Gamma_{F^4/X^4}=\alpha\int d^4x\frac{V^2\bar
V^2}{(\varphi^i\bar\varphi_i)^2}\,.
\ee
The action $S_2+\Gamma_{F^4/X^4}$ produces the following equations
of motion for the auxiliary fields $V_{\alpha\beta}$ and
$\bar V_{\dot\alpha\dot\beta}$,
\be
F_{\alpha\beta}=V_{\alpha\beta}\left[1+\alpha\frac{\bar V^2}{(\varphi^i\bar \varphi_i)^2}
\right],\qquad
\bar F_{\dot\alpha\dot\beta}=\bar V_{\dot\alpha\dot\beta}
 \left[1+\alpha\frac{ V^2}{(\varphi^i\bar \varphi_i)^2}
\right].
\ee
The solution of these equations can be represented as a series
in the Maxwell field strength, in which we need only the lowest
terms,
\be
V_{\alpha\beta}=F_{\alpha\beta}\left[
1-\alpha\frac{\bar F^2}{(\varphi^i\bar \varphi_i)^2}+O(F^3)
\right],\qquad
\bar V_{\dot\alpha\dot\beta}=\bar F_{\dot\alpha\dot\beta}\left[
1-\alpha\frac{F^2}{(\varphi^i\bar \varphi_i)^2}+O(F^3)
\right].
\ee
Substituting these solutions back into $S_2+\Gamma_{F^4/X^4}$, we earn
the correct $F^6$ term,
\be
S_2+\Gamma_{F^4/X^4}=\int d^4x\left[
-\frac12(F^2+\bar F^2)+\alpha\frac{F^2\bar F^2}{(\varphi^i\bar\varphi_i)^2}
-\alpha^2\frac{F^2\bar F^2}{(\varphi^i\bar\varphi_i)^2}(F^2+\bar F^2)
+O(F^8)
\right].
\ee
With $\alpha=Q/2$ this action coincides, up to the $F^6$ order, with
the Born-Infeld part of the effective worldvolume action for a D3 brane moving in curved
$AdS_5\times S^5$ vacuum background of type IIB supergravity
\cite{MT}\footnote{We omit here all terms with derivatives of scalars $X^I$.},
\bea
&&\int d^4x\,\frac{|X|^4}Q\left[1-
\sqrt{-\det(\eta_{mn}+Q^{1/2}|X|^{-2}F_{mn})}
\right]
\nn\\&=&
-\frac12\int d^4x[F^2+\bar F^2-\frac{Q}{|X|^4}F^2\bar F^2
+\frac12\frac{Q^2}{|X|^8}F^2\bar F^2(F^2+\bar F^2)+O(F^8)]
\,.
\eea

In stringy language, we can make the identification $Q= N g_s\alpha'{}^2/\pi$, where $N$ is
the number of D3 branes which induce the $AdS_5\times S^5$
geometry, $g_s$ is the string coupling and $\alpha'$ is the
inverse string tension. It was conjectured in
\cite{CT,M,KK,Alwis,BGL} (see also \cite{BI-review} for a review)
that such D3 brane effective action should coincide with
the low-energy effective action of $\cN=4$ SU($N$) SYM theory in the
large $N$ limit. Thus here we proved this conjecture up to the $F^6/X^8$
order.

\subsection{A comment on the Wess-Zumino term}
\label{WZ-comment}
As shown in \cite{TZ,Intriligator}, the quantum effective action of ${\cal N}=4$ SYM theory must contain a Wess-Zumino (WZ)
type non-tensor term in the scalar fields sector. The presence of such term in various superfield versions of the ${\cal N}=4$ SYM
effective action
(in particular, in its ${\cal N}=2$ superfield version \cite{BuIv}) was recently proved in \cite{BS1,BS2}. Here we give an evidence
that the ${\cal N}=3$ effective action (\ref{G}) also contains WZ term in its component expansion.

To detect the WZ term, it is sufficient to keep only scalar fields in the superfields
(\ref{W-comp}),
\bea
\hat\omega_{23}&=&\phi^1
+2i\theta_2^\alpha\bar\theta^{2\dot\alpha}\partial_{\alpha\dot\alpha}\phi^1
-4i\theta_2^\alpha\bar\theta^{1\dot\alpha}\partial_{\alpha\dot\alpha}\phi^2
-4i\theta_3^\alpha\bar\theta^{1\dot\alpha}\partial_{\alpha\dot\alpha}\phi^3
+8\theta_2^\alpha\theta_3^\beta\bar\theta^{1\dot\alpha}\bar\theta^{2\dot\beta}
\partial_{\alpha\dot\alpha}\partial_{\beta\dot\beta}\phi^3\,,
\nn\\
\hat{\bar \omega}{}^{12}&=&\bar\phi_3
-2i\theta_2^\alpha\bar\theta^{2\dot\alpha}\partial_{\alpha\dot\alpha}\bar\phi_3
+4i\theta_3^\alpha\bar\theta^{1\dot\alpha}\partial_{\alpha\dot\alpha}\bar\phi_1
+4i\theta_3^\alpha\bar\theta^{2\dot\alpha}\partial_{\alpha\dot\alpha}\bar\phi_2
+8\bar\theta^{1\dot\alpha}\bar\theta^{2\dot\beta}
  \theta_2^\alpha\theta_3^\beta
  \partial_{\alpha\dot\alpha}\partial_{\beta\dot\beta}\bar\varphi_1
  \,.\nn\\&&
\eea
We substitute these superfields into the action (\ref{G}) and
integrate over the Grassmann variables, keeping only those terms which contain
four derivatives contracted with the antisymmetric $\varepsilon$-symbol,
\bea
\Gamma_{WZ}&=&-\frac{i\alpha}{8}\varepsilon^{mnpq}\int d^4x du
[\partial_m\phi^2 \partial_n\bar\phi_3\partial_p\bar\phi_2\partial_q\phi^3
+\partial_m\bar\phi_2\partial_n\bar\phi_1\partial_p\phi^2\partial_q\phi^1]
\nn\\&&\times
\sum_{i,j=0}^\infty (-1)^{i+j}\frac{(i+j+2)!(i+1)(j+1)}{i!j!}
(c^3\bar\phi_3)^i(\bar c_1\phi^1)^j
\,.
\label{WZ-series}
\eea
To compare this expression with the standard expression (\ref{WZ}) for WZ
term\footnote{To be precise, we compare
(\ref{WZ-series}) with the WZ action in the
four-dimensional form (\ref{SWZ}).}, it is necessary to compute the harmonic integrals and to
sum the series. Unfortunately, it is very difficult to find the
explicit expression for the integral
\be
\int du\,
u^1_{i_1} \bar u_1^{i'_1}\ldots
u^1_{i_n} \bar u_1^{i'_n}
u^3_{j_1} \bar u_3^{j'_1}\ldots
u^3_{j_m} \bar u_3^{j'_m}u^2_k \bar u_2^{k'}
\ee
in terms of (anti)symmetrized irreducible combinations of
the delta-symbols. Therefore here we restrict ourselves to considering only
the lowest terms in (\ref{WZ-series}), namely,
\be
\Gamma_{WZ}=\frac32i\alpha\varepsilon^{mnpq}\int d^4x du
[\partial_m\phi^2 \partial_n\bar\phi_3\partial_p\bar\phi_2\partial_q\phi^3
+\partial_m\bar\phi_2\partial_n\bar\phi_1\partial_p\phi^2\partial_q\phi^1]
(c^3\bar\phi_3+\bar c_1\phi^1)
+O(\phi^6)
\,.
\label{WZ-leading}
\ee
The corresponding harmonic integral is quite easy to do,
\be
\int du\,u^1_iu^2_ju^3_k \bar u_1^{i'} \bar u_2^{j'} \bar u_3^{k'}
=\frac1{36}\varepsilon_{ijk}\varepsilon^{i'j'k'}
+\frac1{60}\delta_i^{(i'}\delta_j^{j'}\delta_k^{k')}
+\frac1{18}\delta_i^{(i'}\delta_j^{[j')}\delta_k^{k']}
+\frac1{18}\delta_i^{[i'}\delta_j^{(j']}\delta_k^{k')}
\,.
\label{25}
\ee
Then it is straightforward to see that only the first term in the r.h.s. of (\ref{25})
contributes to (\ref{WZ-leading}), while all other terms  either vanish
after contracting the indices or form total derivatives. As a result, (\ref{WZ-leading})
can be rewritten as
\bea
\Gamma_{WZ}&=&\frac{i\alpha}{24}\varepsilon^{mnpq}\int d^4x\,
\varepsilon_{ijk}\varepsilon^{i'j'k'}
[c^i\partial_m\phi^j \partial_n \phi^k
\bar\phi_{i'} \partial_p\bar\phi_{j'}\partial_q\bar\phi_{k'}
\nn\\&&\qquad\qquad
-\,\phi^i \partial_m\phi^j\partial_n\phi^k
\bar c_{i'}\partial_p\bar \phi_{j'}\partial_q\bar\phi_{k'}]
+O(\phi^6)\,.
\eea
This expression coincides with (\ref{25_}) under the choice
\be
\alpha=-\frac1{2\pi^2}\,.
\ee
This proves that the action (\ref{G}) contains the Wess-Zumino
term. One of the possible four-dimensional representations of this term is given
by the expression (\ref{SWZ}).

\section{Summary and discussion}
In the present paper we made an essential step towards solving the
long-standing problem of constructing $\cN=3$ SYM low-energy effective action in
terms of unconstrained $\cN=3$ superfields. We constructed the
leading part of this effective action which is responsible for the
$F^4/X^4$ term in components. This action is given by a
local functional in the $\cN=3$ analytic superspace, such that it depends on
the $\cN=3$ superfield strength without derivatives on them. The form of this functional is
uniquely fixed by the requirements of scale and $\gamma_5$ invariance,
although the action respects further SU(2,2$|$3) superconformal symmetry
(and, perhaps, SU(2,2$|$4)).

Since the $\cN=3$ and $\cN=4$ SYM models are equivalent on shell,
the action (\ref{G}) provides us with an $\cN=3$ superfield
description of the $\cN=4$ SYM low-energy effective action.
This effective action was previously studied in the $\cN=2$
harmonic superspace \cite{BuIv,BIP} and was rewritten in terms of the
on-shell $\cN=4$
superfields in \cite{BLS,BS1,BS2}. In contrast to the representations
of this action in the $\cN=2$ and $\cN=4$ harmonic superspaces,
the Lagrangian in (\ref{G}) has an explicit dependence on the vev
constants $c^i=\langle\varphi^i\rangle$. However, this dependence
is rather spurious: we proved that the action itself is in fact independent of any
particular choice of these constants. This phenomenon is very similar to what one observed
in the action of the $\cN=2$ improved tensor multiplet \cite{deWit} in the harmonic
superspace \cite{GIO} which also explicitly included the vev constants of the scalars,
but this dependence disappeared in the full component action. In \cite{GIO} it was argued that
the presence of such constants reflects the non-trivial topological properties of
this action. In our case the $\cN=4$ SYM low-energy
effective action also contains some topological term given by the
WZ action for the scalar fields \cite{TZ,Intriligator}.
Therefore the action (\ref{G}) can be equally considered as an $\cN=3$
superfield extension of the WZ term.
A possible form of the WZ term arising from the $\cN=3$ harmonic
superspace is discussed in Appendix C, see eq. (\ref{SWZ}).
The constants $c^i$ in this action break the
manifest SU(3) symmetry, though the action is still SU(3) and SU(4)
invariant up to total derivatives. This confirms the conclusions
of \cite{BS1} that the four-dimensional WZ term cannot be made manifestly
invariant under SU(3) since this group is anomalous.

In the present paper we studied the bosonic component structure of the
action (\ref{G}) in the limit of constant Maxwell and scalar
fields and argued that it contains the Wess-Zumino term.
We showed that this action correctly reproduces the coefficients in
front of the $F^4/X^4$ and $F^6/X^8$ terms to ensure their coincidence with the similar
terms in the worldvolume action of D3 brane in the $AdS_5\times S^5$ background. To make the comparison of
the action (\ref{G}) with the D3 brane action more precise, it is necessary to
study the component structure of (\ref{G}) in the scalar field
sector in more detail, beyond the constant field
approximation. This problem is technically involved
and will be addressed elsewhere.

Finally, it is worth pointing out that the action (\ref{G}) was derived
solely by employing the group-theory requirements of gauge invariance and superconformal
symmetry. It is very desirable to develop the background field
method for the $\cN=3$ SYM theory in order to re-derive the action (\ref{G})
from the quantum perturbation theory in $\cN=3$ harmonic superspace\footnote{Like as the ${\cal N}=2$ superfield effective
action of ref.\ \cite{BuIv} was re-derived from the ${\cal N}=2$ harmonic superfield perturbation theory in
\cite{BIP}.}.
Note that the free propagators in the $\cN=3$ harmonic superspace were
studied in \cite{DM}. These methods might help to unveil  the
structure of effective action in the $\cN=3$ and $\cN=4$ SYM
models beyond the low-energy approximation.

\vspace{30pt}
\noindent
{\bf Acknowledgments}\\[3mm]
I.B.S.\ is indebted to D.~Belyaev, S.~Kuzenko, W.~Schulgin and D.~Sorokin for
useful discussions. I.L.B.\ is grateful to CAPES for supporting his visit to the Physics
Department of Universidade Federal de Juiz de Fora where the final
part of work was done.
The authors are grateful to the RFBR grant Nr.\ 11-02-90445 for partial support.
I.L.B.\ and I.B.S.\ acknowledge the support from the RFBR grant Nr.\ 12-02-00121 and
from LRSS grant Nr.\ 224.2012.2.
The work of I.B.S.\ was also supported by the Marie Curie research fellowship Nr.\ 236231,
``QuantumSupersymmetry''. E.A.I. and B.M.Z. acknowledge the support from
the RFBR grant Nr.\ 09-02-01209 and a grant of Heisenberg-Landau Program.
E.A.I.\ thanks the Directorate of SUBATECH, University of Nantes, for the kind hospitality at the final stage
of this work.

\appendix

\section{Derivation of scale and $\gamma_5$ invariant
effective action}
Here we derive the equations for the coefficients
$\alpha_{m,n}$ which follow from the requirement that the action
(\ref{Gn}) is invariant under (\ref{scale}). Consider two lowest
terms in the series (\ref{Gn})\footnote{Here, for simplicity, we put $c^i\bar
c_i=1$ and $\alpha=32\,$.},
\bea
\Gamma_0&=&\alpha_{0,0}\int
d\zeta(^{33}_{11})du(\bar\omega^{12}\omega_{23})^2\,,\nn\\
\Gamma_1&=&\alpha_{0,1}\int d\zeta(_{11}^{33})du
(\bar\omega^{12}\omega_{23})^2(\bar\omega^{12}c^3+\omega_{23}\bar
c_1)\,.
\label{59}
\eea
The superconformal variation of $\Gamma_0$ reads
\be
\delta_{\rm sc}\Gamma_0=2\alpha_{0,0}(A+\bar A)\int d\zeta(^{33}_{11})du
(\bar\omega^{12}\omega_{23})^2\,.
\label{var-59}
\ee
Note that the terms with
$\bar\omega^{12}\bar\omega^{12}\omega_{23}$ and
$\bar\omega^{12}\omega_{23}\omega_{23}$ vanish on shell because of
the relations (\ref{prop}).

The superconformal variation of $\Gamma_1$ reads
\be
\delta_{\rm sc}\Gamma_1=3\alpha_{0,1}\int d\zeta(^{33}_{11})du
\left[(\bar\omega^{12}\omega_{23})^2(\bar A c^3\bar c_3+A c^1 \bar c_1)+
O(\omega^5)\right].
\label{60}
\ee
Using the identities
\be
c^1=D^1_2 c^2= D^1_3 c^3\,,\qquad
\bar c_3 =-D^1_3 \bar c_1=-D^2_3 \bar c_2\,,
\ee
which follow from the definitions (\ref{cccc}), one can write
\bea
c^1\bar c_1&=&\frac13(c^1\bar c_1+\bar c_1 D^1_2 c^2 +\bar c_1 D^1_3
c^3)\,,\nn\\
c^3\bar c_3&=&\frac13(c^3\bar c_3-c^3 D^1_3 \bar c_1- c^3 D^2_3 \bar c_2)\,.
\eea
We substitute these expressions into (\ref{60}) and integrate by
parts with respect to the harmonic derivatives $D^1_2$, $D^2_3$ and $D^1_3$,
\be
\delta_{\rm sc}\Gamma_1=\alpha_{0,1}\int d\zeta(^{33}_{11})du
\left[(\bar A+A)(\bar\omega^{12}\omega_{23})^2+
O(\omega^5)\right].
\label{61}
\ee
Here we made also use of the identity $c^1\bar c_1+c^2\bar c_2+c^3\bar c_3=c^i\bar c_i=1$.
Comparing (\ref{61}) with (\ref{var-59}), we observe that the terms
with four superfield strengths are canceled out under the condition
\be
\alpha_{0,1}=-2\alpha_{0,0}\,.
\ee

Let us now consider the $n$-th term in the series (\ref{Gn}),
\be
\Gamma_n=\int d\zeta(^{33}_{11})du
(\bar \omega^{12}\omega_{23})^2
\sum_{i=0}^{n}
\alpha_{i,n-i}
(\bar\omega^{12}c^3)^i
(\omega_{23}\bar c_1)^{n-i}\,,
\label{Gn_}
\ee
and compute its variation under (\ref{scale}),
\bea
\delta_{\rm sc}\Gamma_n&=&
\int d\zeta(^{33}_{11})du
(\bar \omega^{12}\omega_{23})^2
\sum_{i=0}^{n}
\alpha_{i,n-i}[(i+2)\bar A+(n-i+2)A]
(\bar\omega^{12}c^3)^i
(\omega_{23}\bar c_1)^{n-i}
\nn\\&&
+\int d\zeta(^{33}_{11})du
(\bar \omega^{12}\omega_{23})^2
\sum_{i=1}^{n}
\alpha_{i,n-i}(i+2)\bar A
(\bar\omega^{12}c^3)^{i-1}
(\omega_{23}\bar c_1)^{n-i}c^3\bar c_3
\nn\\&&
+\int d\zeta(^{33}_{11})du
(\bar \omega^{12}\omega_{23})^2
\sum_{i=0}^{n-1}
\alpha_{i,n-i}(n-i+2) A
(\bar\omega^{12}c^3)^i
(\omega_{23}\bar c_1)^{n-i-1}c^1\bar c_1\,.
\label{47}
\eea
In the second line of (\ref{47}) we apply the identity
\be
\bar c_3(c^3)^i(\bar c_1)^{n-i}=
[\frac{i}{n+2}\bar c_3-\frac{n-i+1}{n+2}D^1_3\bar c_1
-\frac1{n+2}D^2_3 \bar c_2](c^3)^i(\bar c_1)^{n-i}\,.
\ee
Upon integrating by parts with respect to the harmonic derivatives $D^1_3$ and
$D^2_3$, this expression is replaced by
\be
\frac i{n+2}(\bar c_1)^{n-i}(c^3)^{i-1}\,.
\ee
Similarly, in the last line of (\ref{47}) we apply the identity
\be
c^1(\bar c_1)^{n-i}(c^3)^i=
[\frac{n-i}{n+2}c^1+\frac1{n+2}D^1_2 c^2+\frac{i+1}{n+2}D^1_3 c^3]
(c_1)^{n-i}(c^3)^i
\ee
and again integrate by parts with respect to the harmonic derivatives. As a result,
the expression $c^1(\bar c_1)^{n-i}(c^3)^i$ in (\ref{47}) produces the term
\be
\frac{n-i}{n+2}(c^3)^i(\bar c_1)^{n-i-1}\,.
\ee

Taking all this into account, the variation (\ref{47}) can be written as
\bea
\delta_{\rm sc}\Gamma_n&=&
\int d\zeta(^{33}_{11})du
(\bar \omega^{12}\omega_{23})^2
\sum_{i=0}^{n}
\alpha_{i,n-i}[(i+2)\bar A+(n-i+2)A]
(\bar\omega^{12}c^3)^i
(\omega_{23}\bar c_1)^{n-i}
\nn\\&&
+\int d\zeta(^{33}_{11})du
(\bar \omega^{12}\omega_{23})^2
\sum_{i=1}^{n}
\alpha_{i,n-i}\frac{i(i+2)}{n+2}\bar A
(\bar\omega^{12}c^3)^{i-1}
(\omega_{23}\bar c_1)^{n-i}
\\&&
+\int d\zeta(^{33}_{11})du
(\bar \omega^{12}\omega_{23})^2
\sum_{i=0}^{n-1}
\alpha_{i,n-i}\frac{(n-i)(n-i+2)}{n+2} A
(\bar\omega^{12}c^3)^i
(\omega_{23}\bar c_1)^{n-i-1}\,.
\nn
\label{48}
\eea
We observe that the terms in the last two lines in (\ref{48})
cancel similar terms in the first line of
$\delta_{\rm sc}\Gamma_{n-1}$, provided the coefficients $\alpha_{ij}$ obey the
following two equations
\bea
\alpha_{i,n-i}\,\dfrac{(n-i+2)(n-i)}{n+2}
+\alpha_{i+1,n-i-1}\,\dfrac{(i+3)(i+1)}{n+2}&=&-(n+3)\alpha_{i,n-i-1}
\,,\\
\alpha_{i,n-i}\,\dfrac{(n-i+2)(n-i)}{n+2}
-\alpha_{i+1,n-i-1}\,\dfrac{(i+3)(i+1)}{n+2}&=&-(n-2i-1)\alpha_{i,n-i-1}\,.
\nn
\eea
As a consequence, any two adjacent coefficients are related as
\be
\frac{\alpha_{i,j}}{\alpha_{i,j-1}}
=-\frac{(j+1)(i+j+2)}{(j+2)j}\,.
\ee
The solution of this equation is just (\ref{alpha}).

\section{Harmonic integrals}
The standard definition of the integration over the SU(3) harmonic
variables reads \cite{hss-book}
\be
\int du\,1=1\,,\qquad
\int du(\mbox{non-singlet SU(3) irreducible representation})=0\,.
\ee
{}From this definition one can derive the following simple relations
\be
\int du\, u^1_i \bar u_1^j=\int du\, u^3_i \bar
u_3^j=\frac13\delta_i^j\,,\quad
\int du\, u^1_i\bar u_1^ju^1_k\bar u_1^l
  =\frac1{6}\delta_i^{(j}\delta_k^{l)}\,,
\quad\mbox{etc.}
\ee
All these integrals appear as particular cases of the following
general formula
\bea
\int du\,
u^1_{i_1} \bar u_1^{i'_1}\ldots
u^1_{i_n} \bar u_1^{i'_n}
u^3_{j_1} \bar u_3^{j'_1}\ldots
u^3_{j_m} \bar u_3^{j'_m}&=&\sum_{k=0}^m
\frac{2m!(-1)^k}{(m+1)(k+n+2)(k+n+1)k!(m-k)!}\nn\\&&
\times\,
\delta_{i_1}^{(i'_1}\ldots \delta_{i_n}^{i'_n}
\delta_{(j_1}^{\{j'_1}\ldots \delta_{j_k}^{j'_k)}\ldots
\delta_{j_m)}^{j'_m\}}\,.
\eea
Here both $(\ldots)$ and $\{\ldots\}$ denote symmetrization of
the indices.
Contracting this expression with vev constants $c^i$, $\bar c_i$
and with scalar fields $\phi^i$, $\bar\phi_i$ we find
\bea
\int du(\phi^1\bar c_1)^n (c^3\bar\phi_3)^m
&=&\sum_{k=0}^m
\frac{2m!(-1)^k}{(m+1)(k+n+2)(k+n+1)k!(m-k)!}\\&&
\times\phi^{(i_1}\ldots \phi^{i_n}c^{j_1}\ldots c^{j_k)}\ldots c^{j_m}
\bar c_{i_1}\ldots \bar c_{i_n}\bar \phi_{j_1}\ldots \bar \phi_{j_k}
\ldots \bar \phi_{j_m}\,.
\nn
\eea
After applying some combinatorics, this expression can be represented
in the following useful form
\bea
\int du(\phi^1\bar c_1)^n (c^3\bar\phi_3)^m
&=&
\sum_{k=0}^{{\rm min}(m,n)} \frac{2n!m!(m+n-k+1)!(-1)^k}{k!(n-k)!(m-k)!(m+n+2)!(n+1)(m+1)}
\nn\\&&\times\,
(\phi^i\bar\phi_i)^k
(\phi^i\bar c_i)^{n-k}(c^i \bar\phi_i)^{m-k}\,.
\label{hint}
\eea

\section{Wess-Zumino term}
\subsection{Derivation from five-dimensions}
Consider six real scalar fields $X^A$, $A=1,\ldots,6$, in the
fundamental representation of SO(6).
The WZ term for these scalars has the standard form
\cite{TZ,Intriligator}
\be
S_{WZ}=-\frac1{60\pi^2}\int d^5x\,
\varepsilon^{MNKLP}\varepsilon^{ABCDEF}\frac1{|X|^6}
X^A\partial_M X^B \partial_N X^C
\partial_K X^D \partial_L X^E \partial_P X^F\,,
\label{WZ}
\ee
where $|X|^2=X^A X^A$. It is useful to introduce the normalized
scalars,
\be
Y^A=\frac{X^A}{|X|}\,,\qquad
Y^A Y^A=1\,,
\ee
in terms of which the action (\ref{WZ}) can be rewritten as
\be
S_{WZ}=-\frac1{60\pi^2}\int d^5x\,
\varepsilon^{MNKLP}\varepsilon^{ABCDEF}
Y^A\partial_M Y^B \partial_N Y^C
\partial_K Y^D \partial_L Y^E \partial_P Y^F\,.
\label{WZ1}
\ee
The integration here is performed over a five-dimensional manifold
$\cal M$ which has the four-dimensional Minkowski space as its
boundary $\partial\cal M$.

Let us rewrite the action (\ref{WZ1}) in the manifestly SU(3) covariant form.
For this purpose we pass to the complex scalars,
\bea
&&
f^1=Y^1+iY^2\,,\quad
f^2=Y^3+iY^4\,,\quad
f^3=Y^5+iY^6\,,\nn\\&&
\bar f_1=Y^1-iY^2\,,\quad
\bar f_2=Y^3-iY^4\,,\quad
\bar f_3=Y^5-iY^6\,,
\label{ff}
\eea
which are also normalized,
\be
f^i \bar f_i =1\,.
\label{norm}
\ee
In terms of these scalars the action (\ref{WZ1}) acquires the desired manifestly SU(3)
covariant form,
\bea
S_{WZ}&=&\frac i{48\pi^2}\varepsilon^{MNKLP}
\varepsilon_{ijk}\varepsilon^{i'j'k'}
\int d^5x[-(f^i\partial_M f^j \partial_N f^k)
\partial_K(\bar f_l \partial_L \bar f_m \partial_P \bar f_n)
\nn\\&&
+\,
\partial_K(f^i\partial_M f^j \partial_N f^k)
(\bar f_{i'} \partial_L \bar f_{j'} \partial_P \bar f_{k'})
]\,.
\label{WZ2}
\eea
It is also useful to rewrite (\ref{WZ2}) as
\be
S_{WZ}=\frac i{48\pi^2}\int_{\cal M} (d\omega_2\wedge \bar\omega_2
-\omega_2\wedge d\bar\omega_2)\,,
\label{WZ3}
\ee
where $\omega_2$ and $\bar\omega_2$ are 2-forms,
\be
\omega_2=\varepsilon_{ijk}f^i df^j\wedge df^k\,,\qquad
\bar\omega_2=\varepsilon^{ijk}\bar f_i d\bar f_j \wedge d\bar
f_k\,.
\ee
Note that the action (\ref{WZ3}) is real.

The equation (\ref{norm}) has the evident consequence
\be
d f^i \bar f_i + f^i d\bar f_i=0\,.
\ee
Using this relation, it is easy to prove the following important
identity
\be
\omega_2\wedge d\bar\omega_2=-d\omega_2\wedge \bar\omega_2\,,
\ee
or
\be
d(\omega_2\wedge\bar\omega_2)=0\,.
\label{10}
\ee
With taking into account this identity, the WZ action (\ref{WZ3}) acquires the form
\be
S_{WZ}=\frac i{24\pi^2}\int_{\cal M} d\omega_2\wedge\bar \omega_2\,.
\label{WZ4}
\ee

Now, let us introduce the projections
\be
y=f^i \bar c_i\,,\qquad
\bar y=\bar f_i c^i\,,
\ee
where $c^i$ are {\it arbitrary} constants with the SU(3) index.
Owing to the identities
\be
dy\wedge\omega_2=\frac y3d\omega_2\,,
\qquad
d\bar y\wedge\bar \omega_2=\frac{\bar y}3d\bar\omega_2\,,
\ee
the WZ action (\ref{WZ3}) can be rewritten as
\be
S_{WZ}=\frac{i}{8\pi^2}\int_{\cal M} \frac 1y dy\wedge
\omega_2\wedge\bar\omega_2
=\frac{i}{8\pi^2}\int_{\cal M} d\ln y \wedge
\omega_2\wedge\bar\omega_2\,,
\ee
or, in the self-conjugated form,
\be
S_{WZ}=\frac{i}{16\pi^2}\int_{\cal M} d\ln\frac{ y}{\bar y} \wedge
\omega_2\wedge\bar\omega_2\,.
\ee
Due to the identity (\ref{10}), it is easy to integrate this
form and to rewrite $S_{WZ}$ as an integral over the $d=4$ Minkowski boundary
\be
S_{WZ}=\frac{i}{16\pi^2}\int_{\cal M}
d[\ln\frac{ y}{\bar y}\, \omega_2\wedge\bar\omega_2]
=\frac{i}{16\pi^2}\int_{\partial\cal M}
\ln\frac{ y}{\bar y}\,
\omega_2\wedge\bar\omega_2+\chi_4\,,
\label{WZ5}
\ee
where $\chi_4$ is an arbitrary closed 4-form, $d\chi_4=0$.
For simplicity, in what follows we assume that $\chi_4=0$, but, in general, there
is no any prescription how to fix this 4-form.

In terms of the scalars (\ref{ff}), the action (\ref{WZ5}) can be cast in the following explicit form,
\be
S_{WZ}=\frac{i}{16\pi^2}
\varepsilon^{mnpq}\varepsilon_{ijk}\varepsilon^{i'j'k'}
\int d^4x\, \ln\frac{f^l \bar c_l}{\bar f_{l'} c^{l'}}
(f^i \partial_m f^j \partial_n f^k)
(\bar f_{i'} \partial_p \bar f_{j'} \partial_q \bar f_{k'})\,.
\ee
It is now easy to come back to the non-normalized scalars,
$f^i=\varphi^i/\sqrt{\varphi^l\bar\varphi_l}$,
$\bar f_i=\bar\varphi_i/\sqrt{\varphi^l\bar\varphi_l}$ and to obtain the
final four-dimensional form of the WZ action:
\be
S_{WZ}=\frac{i}{16\pi^2}
\varepsilon^{mnpq}\varepsilon_{ijk}\varepsilon^{i'j'k'}
\int d^4x\, \ln\frac{\varphi^l \bar c_l}{\bar \varphi_{l'} c^{l'}}
\frac{
(\varphi^i \partial_m \varphi^j \partial_n \varphi^k)
(\bar \varphi_{i'} \partial_p \bar \varphi_{j'} \partial_q \bar \varphi_{k'})}{
(\varphi^i\bar\varphi_i)^3}\,.
\label{SWZ}
\ee
The constants $c^i$ in this action break the explicit SU(3)
invariance. Nevertheless, (\ref{SWZ}) is SU(3) invariant (up to total
derivatives), since it was derived from the SU(3) covariant
five-dimensional action (\ref{WZ2}). The same argument implies that (\ref{SWZ})
respects a hidden SO(6)$\sim$ SU(4) invariance.

\subsection{Expansion around vacuum}
We point out that the constants $c^i$
in (\ref{SWZ}) are arbitrary. In this subsection we show that this
action can appear in the component field expansion of the
superfield action (\ref{G}) if the constants $c^i$ coincide
with the vevs of the scalars.

Let us assume that the constants $c^i$ in
(\ref{SWZ}) are given by (\ref{c}) and make the series expansion
of this action around these vevs,
\bea
\label{21}
S_{WZ}&=&\frac{i}{16\pi^2}
\varepsilon^{mnpq}\varepsilon_{ijk}\varepsilon^{i'j'k'}
\int d^4x
(c^i+\phi^i) \partial_m \phi^j \partial_n \phi^k
(\bar c_{i'} +\bar \phi_{i'}) \partial_p \bar \phi_{j'} \partial_q \bar \phi_{k'}
\\&&\times
\sum_{l=1}^\infty \frac{(-1)^{l+1}}{l}[(\phi^i\bar c_i)^l-(c^i\bar \phi_i)^l]
\frac12\sum_{m,n,k=0}^\infty \frac{(m+n+k+2)!}{m!n!k!}
(c^i\bar\phi_i)^m(\bar c_i\phi^i)^n (\phi^i\bar\phi_i)^k\,.
\nn
\eea
Here the fields $\phi^i$ are related with $\varphi^i$ as in
(\ref{vev-shift}) and we assume that
$c^i \bar c_i=1\,$.

Let us single out in the series (\ref{21}) the terms with minimal
numbers of fields $\phi^i$ and $\bar\phi_i$. These terms
correspond to the choice $m=n=0$ and $l=1$ in the second line in (\ref{21}),
\be
S_{WZ}=\frac{i}{16\pi^2}
\varepsilon^{mnpq}\varepsilon_{ijk}\varepsilon^{i'j'k'}
\int d^4x
(\phi^l\bar c_l-c^l\bar \phi_l)
c^i \partial_m \phi^j \partial_n \phi^k
\bar c_{i'}  \partial_p \bar \phi_{j'} \partial_q \bar
\phi_{k'}+O(\phi^6)\,.
\label{22}
\ee

Up to total derivatives, the following identity holds:
\bea
&&\varepsilon^{mnpq}\varepsilon_{ijk}\varepsilon^{i'j'k'}
(\phi^l\bar c_l-c^l\bar \phi_l)
c^i \partial_m \phi^j \partial_n \phi^k
\bar c_{i'}  \partial_p \bar \phi_{j'} \partial_q \bar
\phi_{k'}
\\&=&
\frac13\varepsilon^{mnpq}\varepsilon_{ijk}\varepsilon^{i'j'k'}
(\phi^i \partial_m \phi^j \partial_n \phi^k
\bar c_{i'}  \partial_p \bar \phi_{j'} \partial_q \bar
\phi_{k'}
-c^i \partial_m \phi^j \partial_n \phi^k
\bar \phi_{i'}  \partial_p \bar \phi_{j'} \partial_q \bar
\phi_{k'})
+\mbox{tot. deriv.}\nn
\eea
Using it, the action (\ref{22}) can be written as
\bea
S_{WZ}&=&\frac{i}{48\pi^2}
\varepsilon^{mnpq}\varepsilon_{ijk}\varepsilon^{i'j'k'}
\int d^4x(
\phi^i \partial_m \phi^j \partial_n \phi^k
\bar c_{i'}  \partial_p \bar \phi_{j'} \partial_q \bar
\phi_{k'}
\nn\\&&\qquad\qquad
-c^i \partial_m \phi^j \partial_n \phi^k
\bar \phi_{i'}  \partial_p \bar \phi_{j'} \partial_q \bar
\phi_{k'})+
O(\phi^6)\,.
\label{25_}
\eea
The assumption that $c^i$ in (\ref{SWZ}) coincide with the vevs was
essential in deriving this expression. In Sect.\ \ref{WZ-comment}
we showed that precisely these terms follow from the $\cN=3$ superfield
action (\ref{G}).

\section{Effective equations of motion}

Effective actions in quantum field
theory can be used to obtain effective equations of motion which
describe the dynamics of fields with taking account of quantum corrections.
For this purpose an effective
action should be well defined off the classical mass shell, but this is not
always possible. For instance, in \cite{BS1,BS2,BLS} the constrained
$\cN=4$ superfields were used for constructing the $\cN=4$ SYM
low-energy effective action. Such on-shell effective actions should be
treated rather in the S-matrix sense, but they cannot be used for
obtaining the effective equations of motion. One of the merits of the $\cN=3$
harmonic superspace approach is the possibility to relax
the on-shell constraints for the superfield strengths and to
express them in terms of unconstrained gauge superfield
potentials. As a result, we are able to derive the
superfield equations of motion which follow from the effective
action (\ref{G}).

Recall that the superfield strengths (\ref{W-anal}) are expressed in terms
on the non-analytic gauge prepotentials $V^2_1$ and $V^3_2$ which,
in turn, are related to the analytic ones by the
zero-curvature equations (\ref{zero}). This is completely
analogous to the $\cN=2$ harmonic superspace approach \cite{hss-book}
in which the chiral superfield strength $W$ has a simple
differential expression in terms of the non-analytic gauge
prepotential $V^{--}$ which, in turn, is related to the analytic gauge
potential $V^{++}$ via the zero-curvature equation
$D^{++}V^{--}=D^{--}V^{++}$. The solution of this equation is
known to be non-local with respect to the harmonic variables
\cite{Zupnik87,Zupnik86} and to involve some harmonic
distributions. In our case the solutions of (\ref{zero}) are also
non-local in the harmonic variables, but we avoid using the
harmonic distributions by representing the solutions of
(\ref{zero}) in the pseudo-differential form,
\be
V^2_1=D^2_1D^2_1\frac1{2+D^2_1D^1_2}V^1_2\,,\qquad
V^3_2=D^3_2D^3_2\frac1{2+D^3_2D^2_3}V^2_3\,,
\label{sol_1}
\ee
where
\be
\dfrac1{2+D^2_1D^1_2}=\frac12\sum_{n=0}^\infty(-\frac12D^2_1D^1_2)^n,\qquad
\dfrac1{2+D^3_2D^2_3}=\frac12\sum_{n=0}^\infty(-\frac12D^3_2D^2_3)^n\,.
\label{sol_2}
\ee
The formal expressions (\ref{sol_1}) can be verified to
obey (\ref{zero}) by making use of the commutation relations between the
harmonic derivatives,
\be
[D^1_2,D^2_1]=S_1\,,\qquad
[D^2_3,D^3_2]=S_2\,,
\ee
where $S_1$ and $S_2$ are commuting U(1) generators in the su(3) algebra of harmonic derivatives.
The gauge potentials have the following charges with respect to
these U(1) generators,
\be
S_1 V^1_2 =2 V^1_2\,,\qquad
S_2 V^2_3= 2 V^2_3\,.
\ee
Combining (\ref{sol_1}) with (\ref{W-anal}), we obtain formal
expressions of the superfield strengths in terms on the analytic
gauge potentials,
\be
\bar W^{12}=-\frac14(D^1)^2 D^2_1 D^2_1
\frac1{2+D^2_1 D^1_2}V^1_2\,,\qquad
W_{23}=\frac14(\bar D_3)^2 D^3_2 D^3_2\frac1{2+D^3_2D^2_3}V^2_3\,.
\label{WV}
\ee

Using the equation (\ref{relH}) for the function $H$, one can
easily compute the variation of the action (\ref{G}),
\be
\delta\Gamma=\frac\alpha{8}\int d\zeta(^{33}_{11})du\frac{\delta \bar \omega^{12}\bar\omega^{12}
\omega_{23}\omega_{23}}{(1+\bar \omega^{12}c^3)(1
+\bar \omega^{12}c^3 +\omega_{23} \bar c_1)^2}+{\rm c.c.}\,.
\ee
Here we assume $c^i\bar c_i=1$  for simplicity. Owing to (\ref{WV}), the variation of
the superfield strength $\bar\omega^{12}$ can be expressed through the variation of the
analytic gauge potential $V^1_2$ ,
\be
\delta \bar\omega^{12}=-\frac14(D^1)^2 D^2_1 D^2_1
\frac1{2+D^2_1 D^1_2}\delta V^1_2\,.
\ee
Then the equation of motion produced by the variation of the action (\ref{G})
with respect to $V^1_2$ reads
\be
\frac{\delta\Gamma}{\delta V^1_2}
=-\frac\alpha{32}\frac1{2+D^1_2 D^2_1}D^2_1 D^2_1 (D^1)^2
\frac{\bar\omega^{12} \omega_{23}\omega_{23}}{
(1+\bar \omega^{12}c^3)(1+\bar \omega^{12}c^3+ \omega_{23} \bar
c_1)^2}\,.
\label{eom}
\ee
Applying the tilde-conjugation to (\ref{eom}), one can obtain the
second equation of motion $\delta\Gamma/\delta V^2_3$.

Note that the equation (\ref{eom}) and its conjugate should not be considered separately, but they
should be added to the classical equations of motion associated with the original
Chern-Simons type action of the $\cN=3$ gauge theory \cite{GIKOS1,GIKOS2}.
Together they form the effective equations of motion up to the four-derivative order.



\begin{thebibliography}{99}

\bibitem{BK-book}
  I.~L.~Buchbinder and S.~M.~Kuzenko,
  {\it Ideas and methods of supersymmetry and supergravity: Or a walk through
  superspace}, Bristol, UK: IOP (1998) 656 p.

\bibitem{hss-book}
  A.~S.~Galperin, E.~A.~Ivanov, V.~I.~Ogievetsky and E.~S.~Sokatchev,
  {\it Harmonic superspace},
  Cambridge, UK: Univ. Pr. (2001) 306 p.

\bibitem{GIKOS1}
  A.~Galperin, E.~Ivanov, S.~Kalitzin, V.~Ogievetsky and E.~Sokatchev,
  {\it N=3 Supersymmetric gauge theory}, Phys.\ Lett.\  B {\bf 151} (1985) 215.


\bibitem{GIKOS2}
  A.~Galperin, E.~Ivanov, S.~Kalitzin, V.~Ogievetsky and E.~Sokatchev,
  {\it Unconstrained off-shell N=3 supersymmetric Yang-Mills theory},
  Class.\ Quant.\ Grav.\  {\bf 2} (1985) 155.

\bibitem{DM}
  F.~Delduc and J.~McCabe,
  {\it The quantization of N=3 super Yang-Mills off-shell in harmonic superspace},
  Class.\ Quant.\ Grav.\  {\bf 6} (1989) 233.

\bibitem{IZ}
  E.~A.~Ivanov and B.~M.~Zupnik,
  {\it N=3 supersymmetric Born-Infeld theory},
  Nucl.\ Phys.\  B {\bf 618} (2001) 3,
  {\tt arXiv:hep-th/0110074}.

\bibitem{BISZ}
  I.~L.~Buchbinder, E.~A.~Ivanov, I.~B.~Samsonov and B.~M.~Zupnik,
  {\it Scale invariant low-energy effective action in N=3 SYM theory},
  Nucl.\ Phys.\  B {\bf 689} (2004) 91,
  {\tt arXiv:hep-th/0403053}.

\bibitem{DS}
  M.~Dine and N.~Seiberg,
  {\it Comments on higher derivative operators in some SUSY field theories},
  Phys.\ Lett.\  B {\bf 409} (1997) 239,
  {\tt arXiv:hep-th/9705057}.

\bibitem{S}
  N.~Seiberg,
  {\it Notes on theories with 16 supercharges},
  Nucl.\ Phys.\ Proc.\ Suppl.\  {\bf 67} (1998) 158,
  {\tt arXiv:hep-th/9705117}.

\bibitem{deWit:1996}
  B.~de Wit, M.~T.~Grisaru and M.~Ro\v{c}ek,
  {\it Nonholomorphic corrections to the one-loop N=2 super Yang-Mills action},
  Phys.\ Lett.\  B {\bf 374} (1996) 297,
  {\tt arXiv:hep-th/9601115}.

\bibitem{non-hol1}
  V.~Periwal and R.~von Unge,
  {\it Accelerating D-branes},
  Phys.\ Lett.\  B {\bf 430} (1998) 71,
  {\tt arXiv:hep-th/9801121}.

\bibitem{non-hol2}
  F.~Gonzalez-Rey and M.~Ro\v{c}ek,
  {\it Nonholomorphic N=2 terms in N=4 SYM: 1-loop calculation in N=2
  superspace},
  Phys.\ Lett.\  B {\bf 434} (1998) 303,
  {\tt arXiv:hep-th/9804010}.

\bibitem{non-hol3}
  I.~L.~Buchbinder and S.~M.~Kuzenko,
  {\it Comments on the background field method in harmonic superspace:
  Non-holomorphic corrections in N=4 SYM},
  Mod.\ Phys.\ Lett.\  A {\bf 13} (1998) 1623,
  {\tt arXiv:hep-th/9804168}.

\bibitem{non-hol4}
  F.~Gonzalez-Rey, B.~Kulik, I.~Y.~Park and M.~Ro\v{c}ek,
  {\it Self-dual effective action of N=4 super-Yang-Mills},
  Nucl.\ Phys.\  B {\bf 544} (1999) 218,
  {\tt arXiv:hep-th/9810152}.

\bibitem{non-hol5}
  E.~I.~Buchbinder, I.~L.~Buchbinder and S.~M.~Kuzenko,
  {\it Non-holomorphic effective potential in N=4 SU(n) SYM},
  Phys.\ Lett.\  B {\bf 446} (1999) 216,
  {\tt arXiv:hep-th/9810239}.

\bibitem{review}
  E.~I.~Buchbinder, B.~A.~Ovrut, I.~L.~Buchbinder, E.~A.~Ivanov, S.~M.~Kuzenko,
  {\it Low-energy effective action in N=2 supersymmetric field theories},
  Phys.\ Part.\ Nucl.\  {\bf 32 } (2001)  641-674.

\bibitem{BuIv}
  I.~L.~Buchbinder and E.~A.~Ivanov,
  {\it Complete N=4 structure of low-energy effective action in N=4
  superYang-Mills theories},
  Phys.\ Lett.\  B {\bf 524} (2002) 208,
  {\tt arXiv:hep-th/0111062}.

\bibitem{BIP}
  I.~L.~Buchbinder, E.~A.~Ivanov and A.~Y.~Petrov,
  {\it Complete low-energy effective action in N=4 SYM: A Direct N=2 supergraph
  calculation},
  Nucl.\ Phys.\  B {\bf 653} (2003) 64,
  {\tt arXiv:hep-th/0210241}.

\bibitem{BBP}
  A.~T.~Banin, I.~L.~Buchbinder and N.~G.~Pletnev,
  {\it One loop effective action for N=4 SYM theory in the hypermultiplet sector:
  Leading low-energy approximation and beyond},
  Phys.\ Rev.\  D {\bf 68} (2003) 065024,
  {\tt arXiv:hep-th/0304046}.

\bibitem{BuPl}
  I.~L.~Buchbinder and N.~G.~Pletnev,
  {\it Hypermultiplet dependence of one-loop effective action in the N=2
  superconformal theories},
  JHEP {\bf 0704} (2007) 096,
  {\tt arXiv:hep-th/0611145}.

\bibitem{GIO}
  A.~Galperin, E.~Ivanov and V.~Ogievetsky,
  {\it Superspace actions and duality transformations for N=2 tensor multiplets},
  Sov.\ J.\ Nucl.\ Phys.\  {\bf 45} (1987) 157
  [Yad.\ Fiz.\  {\bf 45} (1987) 245]
  [Phys.\ Scripta {\bf T15} (1987) 176].

\bibitem{TZ}
  A.~A.~Tseytlin and K.~Zarembo,
  {\it Magnetic interactions of D-branes and Wess-Zumino terms in super
  Yang-Mills effective actions},
  Phys.\ Lett.\  B {\bf 474} (2000) 95, {\tt arXiv:hep-th/9911246}.

\bibitem{Intriligator}
  K.~A.~Intriligator,
  {\it Anomaly matching and a Hopf-Wess-Zumino term in 6d, N=(2,0) field
  theories},
  Nucl.\ Phys.\  B {\bf 581} (2000) 257, {\tt arXiv:hep-th/0001205}.

\bibitem{AFSZ}
  L.~Andrianopoli, S.~Ferrara, E.~Sokatchev and B.~Zupnik,
  {\it Shortening of primary operators in N extended SCFT(4) and harmonic
  superspace analyticity},
  Adv.\ Theor.\ Math.\ Phys.\  {\bf 4} (2000) 1149,
  {\tt arXiv:hep-th/9912007}.

\bibitem{HST}
  P.~S.~Howe, K.~S.~Stelle and P.~K.~Townsend,
  {\it Supercurrents},
  Nucl.\ Phys.\  B {\bf 192} (1981) 332.

\bibitem{BS08}
  I.~L.~Buchbinder and I.~B.~Samsonov,
  {\it N=3 Superparticle model},
  Nucl.\ Phys.\  B {\bf 802} (2008) 180,
  {\tt arXiv:0801.4907 [hep-th]}.

\bibitem{GIO-N3} A.~S.~Galperin, E.~A.~Ivanov, V.~I.~Ogievetsky,
 {\it Superspaces for N=3 supersymmetry}, Sov. J. Nucl. Phys. {\bf 46} (1987)
 543-556.

\bibitem{MT}
  R.~R.~Metsaev and A.~A.~Tseytlin,
  {\it Supersymmetric D3-brane action in $AdS_\times S^5$},
  Phys.\ Lett.\  B {\bf 436} (1998) 281,
  {\tt arXiv:hep-th/9806095}.

\bibitem{CT}
  I.~Chepelev and A.~A.~Tseytlin,
  {\it Long distance interactions of branes: Correspondence between supergravity
  and superYang-Mills descriptions},
  Nucl.\ Phys.\  B {\bf 515} (1998) 73,
  {\tt arXiv:hep-th/9709087}.

\bibitem{M}
  J.~M.~Maldacena,
  {\it The Large N limit of superconformal field theories and supergravity},
  Adv.\ Theor.\ Math.\ Phys.\  {\bf 2} (1998) 231,
  {\tt arXiv:hep-th/9711200}.

\bibitem{KK}
  E.~Keski-Vakkuri and P.~Kraus,
  {\it Born-Infeld actions from matrix theory},
  Nucl.\ Phys.\  B {\bf 518} (1998) 212,
  {\tt arXiv:hep-th/9709122}.

\bibitem{Alwis}
  S.~P.~de Alwis,
  {\it Matrix models and string world sheet duality},
  Phys.\ Lett.\  B {\bf 423} (1998) 59,
  {\tt arXiv:hep-th/9710219}.

\bibitem{BGL}
  V.~Balasubramanian, R.~Gopakumar and F.~Larsen,
  {\it Gauge theory, geometry and the large N limit},
  Nucl.\ Phys.\  B {\bf 526} (1998) 415,
  {\tt arXiv:hep-th/9712077}.

\bibitem{BI-review}
  A.~A.~Tseytlin,
  {\it Born-Infeld action, supersymmetry and string theory},
  In *Shifman, M.A. (ed.): The many faces of the superworld*
  417-452,
  {\tt arXiv:hep-th/9908105}.

\bibitem{BS1} D.~V.~Belyaev and I.~B.~Samsonov,
  {\it Wess-Zumino term in the N=4 SYM effective action
  revisited}, JHEP {\bf 1104} (2011) 112, {\tt arXiv:1103.5070
  [hep-th]}.

\bibitem{BS2}
  D.~V.~Belyaev and I.~B.~Samsonov,
  {\it Bi-harmonic superspace for N=4 d=4 super Yang-Mills},
  JHEP {\bf 1109} (2011) 056, {\tt arXiv:1106.0611 [hep-th]}.

\bibitem{deWit}
  B.~de Wit, R.~Philippe and A.~Van Proeyen,
  {\it The improved tensor multiplet in N=2 supergravity},
  Nucl.\ Phys.\  B {\bf 219} (1983) 143.

\bibitem{BLS}
  I.~L.~Buchbinder, O.~Lechtenfeld and I.~B.~Samsonov,
  {\it N=4 superparticle and super Yang-Mills theory in USp(4) harmonic
  superspace},
  Nucl.\ Phys.\  B {\bf 802} (2008) 208, {\tt arXiv:0804.3063 [hep-th]}.

\bibitem{Zupnik87}
  B.~M.~Zupnik,
  {\it The action of the supersymmetric N=2 gauge theory in harmonic superspace},
  Phys.\ Lett.\  B {\bf 183} (1987) 175.

\bibitem{Zupnik86}
  B.~M.~Zupnik,
  {\it Six-dimensional supergauge theories in the harmonic superspace},
  Sov.\ J.\ Nucl.\ Phys.\  {\bf 44} (1986) 512
  [Yad.\ Fiz.\  {\bf 44} (1986) 794].

\end{thebibliography}
\end{document}